\documentclass[journal, 12pt, draftclsnofoot, onecolumn, twoside]{IEEEtran}

\usepackage{scalerel}
\usepackage{tikz}
\usetikzlibrary{svg.path}

\definecolor{orcidlogocol}{HTML}{A6CE39}
\tikzset{
	orcidlogo/.pic={
		\fill[orcidlogocol] svg{M256,128c0,70.7-57.3,128-128,128C57.3,256,0,198.7,0,128C0,57.3,57.3,0,128,0C198.7,0,256,57.3,256,128z};
		\fill[white] svg{M86.3,186.2H70.9V79.1h15.4v48.4V186.2z}
		svg{M108.9,79.1h41.6c39.6,0,57,28.3,57,53.6c0,27.5-21.5,53.6-56.8,53.6h-41.8V79.1z M124.3,172.4h24.5c34.9,0,42.9-26.5,42.9-39.7c0-21.5-13.7-39.7-43.7-39.7h-23.7V172.4z}
		svg{M88.7,56.8c0,5.5-4.5,10.1-10.1,10.1c-5.6,0-10.1-4.6-10.1-10.1c0-5.6,4.5-10.1,10.1-10.1C84.2,46.7,88.7,51.3,88.7,56.8z};
	}
}

\newcommand\orcidicon[1]{\href{https://orcid.org/#1}{\mbox{\scalerel*{
				\begin{tikzpicture}[yscale=-1,transform shape]
				\pic{orcidlogo};
				\end{tikzpicture}
			}{|}}}}

\usepackage{hyperref}

\ifCLASSINFOpdf
\else
\fi

\usepackage{mathtools}
\usepackage{amssymb}
\usepackage{lmodern}
\usepackage{newtxtext}
\normalfont

\usepackage[T1]{fontenc}
\usepackage{caption}
\usepackage{subcaption}

\usepackage{wrapfig}
\usepackage{url}
\usepackage{mathrsfs}
\usepackage{bbm}
\usepackage{hyperref}
\usepackage{xcolor}
\usepackage{soul}
\usepackage{setspace}
\allowdisplaybreaks
\usepackage{float}
\usepackage[inline]{enumitem} 
\usepackage{soul}
\usepackage{xcolor}

\makeatletter
\newcommand{\@giventhatstar}[2]{#1\,\middle|\,#2}
\newcommand{\@giventhatnostar}[3][]{#1(#2\,#1|\,#3#1)}
\newcommand{\giventhat}{\@ifstar\@giventhatstar\@giventhatnostar}
\makeatother

\newcommand{\E}{\mathbbm{E}}
\makeatletter
\def\endthebibliography{%
	\def\@noitemerr{\@latex@warning{Empty `thebibliography' environment}}%
	\endlist
}
\makeatother

\begin{document}
\title{{Signal Acquisition with Photon-Counting Detector Arrays in Free-Space  Optical Communications}}
\author{Muhammad~Salman~Bashir,~\IEEEmembership{Member,~IEEE,}
	and~Mohamed-Slim~Alouini,~\IEEEmembership{Fellow,~IEEE}
	% <-this % stops a space
	\thanks{This work is supported by Office of Sponsored Research (OSR) at King Abdullah University of Science and Technology (KAUST). \newline M.~S.~Bashir and M.~-S.~Alouini  are with the King Abdullah University of Science and Technology (KAUST), Thuwal 23955-6900, Kingdom of Saudi Arabia.  e-mail: (muhammad.bashir@fulbrightmail.org, slim.alouini@kaust.edu.sa).}% <-this % stops a space
	\thanks{}% <-this % stops a space
	\thanks{}}

\markboth{}%
{}

\markboth{}%
{}

	\maketitle
	
	\begin{abstract}
Pointing and acquisition are an important aspect of free-space optical communications because of the narrow beamwidth associated with the optical signal. In this paper, we have analyzed the pointing and acquisition problem in free-space optical communications for  photon-counting detector arrays and Gaussian beams. In this regard, we have considered the maximum likelihood detection for detecting the location of the array, and analyzed the one-shot probabilities of missed detection and false alarm using the scaled Poisson approximation. Moreover, the upper/lower bounds on the probabilities of missed detection and false alarm for one complete scan are also derived, and these probabilities are compared with Monte Carlo approximations for a few cases. Additionally, the upper bounds on the acquisition time and the mean acquisition time are also derived. The upper bound on mean acquisition time is  minimized numerically with respect to the beam radius for a constant signal-to-noise ratio scenario. Finally, the complementary distribution function of an upper bound on  acquisition time is also calculated in a closed form. Our study concludes that an array of smaller detectors gives a better acquisition performance (in terms of acquisition time) as compared to one large detector of similar dimensions as the array. 
	\end{abstract}
	\IEEEpeerreviewmaketitle
	
	\section{Introduction}
	 Acquisition and tracking is an important problem in the design of free-space optical (FSO) communication systems, especially in the case of  \emph{orbital-angular-momentum} beams \cite{Li}. Moreover, the recent interest in free-space optics for communications between drones and balloons for the Facebook Connectivity and Google's Loon projects \cite{Kaymak}, respectively, has highlighted the importance of acquisition and tracking for minimizing the pointing/geometric loss in FSO communications.
	
	{\emph{Acquisition} is the process whereby a pair of optical terminals obtain each others orientation in space in order to set up a communication link.} Acquisition and tracking systems are needed for the alignment of FSO based transmitters and receivers because of the narrow divergence angles associated with the smaller wavelengths of light: $\theta_{\text{div}}\approx \lambda/D$, where $\theta_{\text{div}}$ is the beam divergence angle, $\lambda$ is the wavelength of the signal, and $D$ is the transmit telescope aperture.  A typical FSO system comprises a laser transmitter, and a photodetector or an array of photodetectors as the front end of the receiver. Noncoherent \emph{intensity modulated direct detection} (IM/DD) schemes are commonly preferred in optical communications receivers for the ease of a simpler implementation. 
	
	{Photon-counting detector arrays are commonly used in deep space optical communications in order to minimize the geometric and pointing loss. In this paper, we have analyzed the acquisition performance for an FSO communication system that utilizes an array of smaller detectors as opposed to one large detector (same dimensions as the array) at the receiver, and  we discovered that  an array of smaller detectors gives a better acquisition performance as compared to a large detector. The detectors employed are the photon-counting type which detect the presence of a signal based on counting photons during an observation interval.} 
	
	{In this study, it is assumed that the beam footprint is smaller than the dimensions of the array. Such a situation arises in the downlink of satellite/drone-ground optical communications because large detector arrays (large compared to the beam footprint) are easier to set up on the ground.  Additionally, the current study also applies to inter-satellite and inter-drone communications when link distances are small so that the assumption of smaller beam footprint on the array holds. For example, for a beam waist (smallest beam radius at the transmitter) of 0.1 meters, the beam footprint/diameter  at a distance of 10 kilometers is approximately 0.2 meters for a wavelength of 1550 nanometers\footnote{This is computed from $\rho(z) \triangleq \rho_0 \sqrt{1 + \left(\frac{\lambda z}{\pi \rho_0^2} \right)^2}$, where $\rho_0$ is known as the beam waist,  $\rho(z)$ is beam radius at distance $z$ meters from the transmitter, and $\lambda$ is wavelength in meters.}. Hence, in a turbulence free environment, using a telescope with a lens of diameter 0.2 meters, or larger, will suffice to capture and focus the transmitted energy onto an array of detectors. However, a lens of diameter larger than 0.2 meters will be required in order to accommodate some beam wander about the mean position.}
		
	{In the following section, we discuss the state-of-the-art acquisition systems for FSO and the contributions of the study in this article.} 
	
	\subsection{Contributions and Organization of this Paper}

	{The major contribution of this paper is the mathematical analysis carried out in order to derive the mean/complementary distribution of an upper bound on acquisition time, and the optimization of the mean acquisition time as a function of beam radius. This analysis  includes the possibility of the requirement of multiple scans of the uncertainty region. This is especially true of low photon rate/signal-to-noise ratio scenario in which the receiver may not be detected during the first scan of the acquisition phase. Additionally, it is also shown that an array of detectors gives a better acquisition performance (in terms of probability of missed detection/false alarm and acquisition time) than a single detector of the same dimensions. 
	 This study is done for a Gaussian beam and  uniform noise (background radiation and thermal noise) case.}

	This paper is organized as follows. Section~\ref{lr} covers the literature review, and Section~\ref{pcd} considers why photon-counting detectors are important and discusses the statistical modeling of the photon counting process with a Poisson model.  Section~\ref{acq_prelim} analyzes the scanning method and other preliminaries, and in Section~\ref{acq_process}, the acquisition performance is analyzed in terms of the probabilities of missed detection and false alarm. Additionally, the effect of uncertainty in beam parameters on the said probabilities is also examined. Furthermore, the complementary cumulative distribution function of an upper bound on the acquisition time is derived in closed form, and is plotted for various detector arrays as a function of noise power and beam radius. The optimization of the mean acquisition time as a function of beam radius is also carried out in the same section. Section~\ref{complexity} examines a brief complexity analysis of acquisition with detector arrays. The conclusions are finalized in Section~\ref{conc}.

	\section{Literature Review} \label{lr}
	The authors in \cite{Ansari} and \cite{Zedini} discuss the performance of the free-space optical communication for channels with pointing errors. Moreover,  \cite{Quwaiee} gives a detailed treatment on the channel capacity for a generalized pointing error model.  Furthermore, the authors in \cite{issaid_TWC_2017} propose a simulation approach in order to ascertain the outage probability of the single and multi-hop FSO links subject to generalized pointing errors. However, in our opinion, there is a shortage of research that treats the acquisition problem in free-space optics from a signal processing  point of view. With the exception of some work discussed in the next paragraphs, majority of the papers on acquisition deal with the hardware solutions for improving the acquisition process. 
	
	The main body of work that deals with acquisition in free-space optics from a signal processing perspective comprises \cite{Wang} and \cite{XinLi}. The authors in \cite{Wang} discuss a two-stage acquisition method for mobile platforms for an array of detectors. During the first stage, the terminals acquire each others location, and during the second, they verify each other's identity through a IV code \cite{Wang}. The focus of the said paper is on achieving secure acquisition between the two mobile communicating terminals.  {However, some important approximation are made in this paper: The detectors of the array with currents above a certain threshold are assigned a unit weight, and the rest are assigned zero weight. Thus, effectively, even though the beam is considered Gaussian, the energy is assumed to be distributed uniformly over the beam footprint. Moreover, even though the acquisition time is optimized with respect to the beam radius, the possibility of multiple uncertainty region scans is not taken into account.} The authors in \cite{XinLi} optimize the mean acquisition time as a function of the uncertainty sphere angle. However, the expressions of probability of acquisition  are not derived. Moreover, the study does not account for detector arrays, and the effect of  beam energy distribution and noise power is not considered either.
	
	Even though the authors in \cite{Farid} and \cite{Mai} discuss pointing errors and do not directly address acquisition, their work is fundamental enough (at least from the pointing perspective) in order to merit discussion in this section. The authors in \cite{Farid} explore the outage capacity optimization in presence of pointing errors. They develop the statistics of the pointing error for circularly shaped detectors and  Gaussian beams, and optimize the outage capacity with respect to the beam radius. The work \cite{Mai} is similar in essence to \cite{Farid}. The authors in \cite{Mai} discuss the maximization of the link availability as a function of the beam radius for fixed beam power, and also explore the the minimization of the transmitted power for the same link availability by tuning to the optimum beam radius.

	There is another body of work that deals with pointing, acquisition and tracking such as \cite{Kaymak}, \cite{Deng}, \cite{Kim}, \cite{Rzasa} and \cite{Kaushal1}. The survey carried out by the authors in \cite{Kaymak} is comprehensive, but the discussion focuses mainly on the hardware, such as the gimbal and mirror based steering PAT systems for tracking  the beam position. The authors in \cite{Deng} deal with the tracking issue by using the output of the camera sensors in order to direct the movement of control moment gyroscopes (CMG) for the \emph{bifocal relay mirror spacecraft} with the help of a feedback loop. The main focus of this work is to suppress the jitter and vibrations in the beam position using CMG's and fine tracking using fast steering mirrors. The work \cite{Kim} proposes gimbal less MEMS micro-mirrors for fast tracking of the time-varying beam position. The authors in \cite{Rzasa} explore the acquisition performance of a gimbal based pointing system experimentally that uses spiral techniques for scanning the uncertainty region. Finally, the paper \cite{Kaushal1} carries out an experimental study in order to find an optimal beam radius that minimizes the bit error rate for a terrestrial channel that undergoes turbulence/scintillation effects.
	
 %The paper is organized as follows. In the next few sections, we touch upon some preliminary, but important, material regarding the benefits of photon counting detector arrays, and the intensity/photon count model for such detector arrays. Section~\ref{acq_prelim} discusses the preliminaries of the acquisition process such as the scanning method of the uncertainty sphere. Section~\ref{acq_process} reviews the acquisition performance in terms of probability of missed detection and false alarm. In the same section, the optimization of the upper bound on mean acquisition time in terms of the beam radius, and the derivation of the complementary distribution function of an upper bound on the acquisition time is also analyzed. The same section also numerically evaluates the performance of the maximum likelihood detection for uncertain beam parameters. Finally, Section~\ref{conc} summarizes the entire study briefly. 
	
	\section{Photon Counting Detector Arrays} \label{pcd}
	\subsection{Why Photon Counting Detectors?} \label{photon_count}
	\begin{figure} 
		\centering
		\includegraphics[scale=0.8]{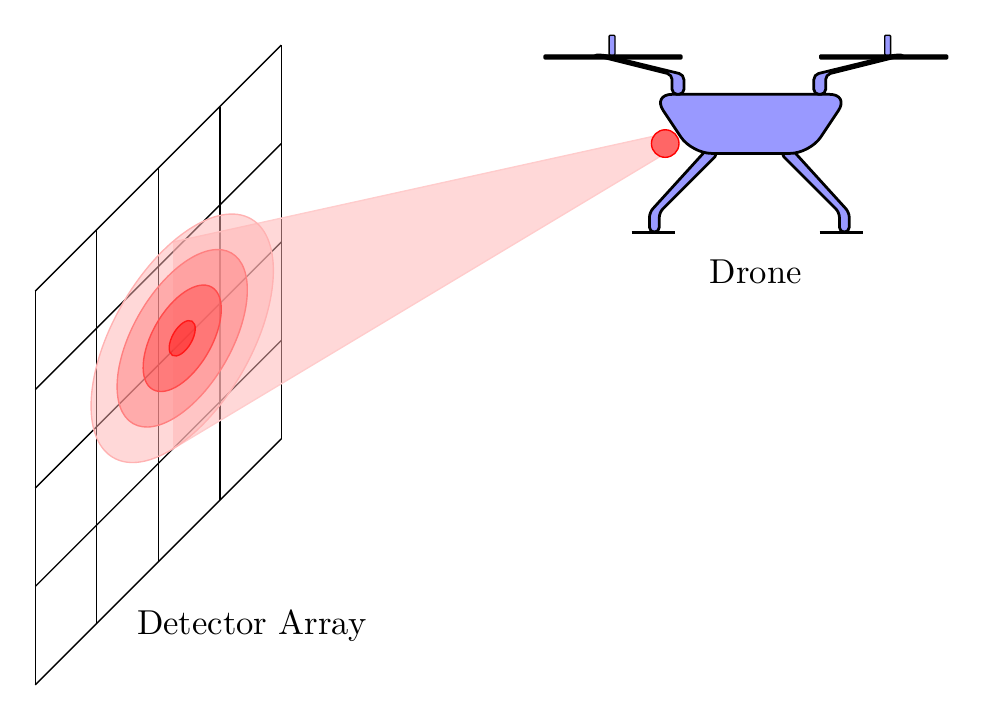}
		\caption{A drone projecting the Gaussian beam on a $4 \times 4$ detector array for the purpose of communication.} \label{drone}
	\end{figure}

%\begin{wrapfigure}{r}{0.5\textwidth}
%	\begin{center}
%		\includegraphics[width=0.48\textwidth]{gauss_beam_on_detector_paper7}
%	\end{center}
%	\caption{A gull}
%\end{wrapfigure}

	Photon counting detectors are preferred over the analog detectors for the better signal-to-noise ratio they provide, especially for low intensity and high frequency optical pulses \cite{Cova}. During the photoamplification process, the photoelectrons emitted from the photodetector surface by the energy of the incident light and background radiation knock electrons out from the cathode of the photomultiplier tube. However, additional noise is further added in the dynode chain after the cathode in the photomultiplier tube. The photon counting detectors can use a \emph{discriminator} to remove the effect of this additional noise which is a less spiky signal as compared to the stream of electrons generated from the cathode. This discrimination process effectively improves the signal-to-noise ratio before the signal is detected. On the other hand, the analog detectors simply ``adds'' the current from the cathode as well as the current (noise) generated after the cathode in the photomultiplier tube, resulting in a lower signal-to-noise ratio. However, analog detectors are simple devices that are cheaper and easier to implement.
	
	\subsection{Communications with an Array of Detectors} \label{array}
	If the intensity profile of the beam is known (or approximately known) at the receiver, an array of smaller detectors is more useful in terms of minimizing the probability of error as compared to one large detector which has the same size as the entire array \cite{Bashir4}. We can think of the communication system with an array of detectors as a \emph{single input multiple output} (SIMO) system. With an array of detectors, the difference in performance in terms of probability of error is more pronounced for low signal-to-noise ratio case. However, even though the probability of error goes down monotonically as the number of detectors go up in the array (the area of the array being constant), the difference in improvement in performance shrinks as the number of detectors becomes large. Hence, we have to settle for a trade-off between the error probability and the number of detectors, since a large number of detectors may not justify additional computational complexity incurred with extra detector circuits. Last, but not the least, a detector array is also more favorable in terms of beam tracking performance \cite{Bashir2}, \cite{Bashir3}, and the study carried out in this paper emphasizes the improvement in acquisition performance with an array of detectors. 
	\subsection{Intensity Profile and the Photodetection Model} \label{profile}

\begin{figure}
	\centering
	\begin{subfigure}[b]{0.32\textwidth}
		\includegraphics[width=\textwidth]{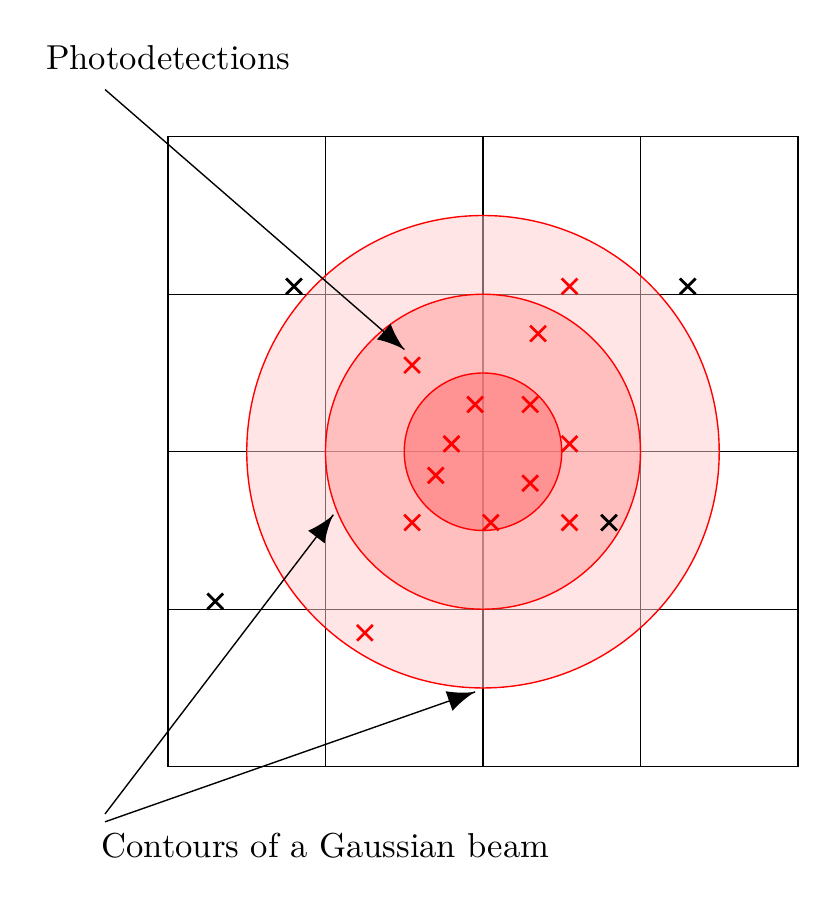}
		\caption{An array of detectors}
		\label{fig11}
	\end{subfigure}
	\begin{subfigure}[b]{0.45\textwidth}
		\includegraphics[width=\textwidth]{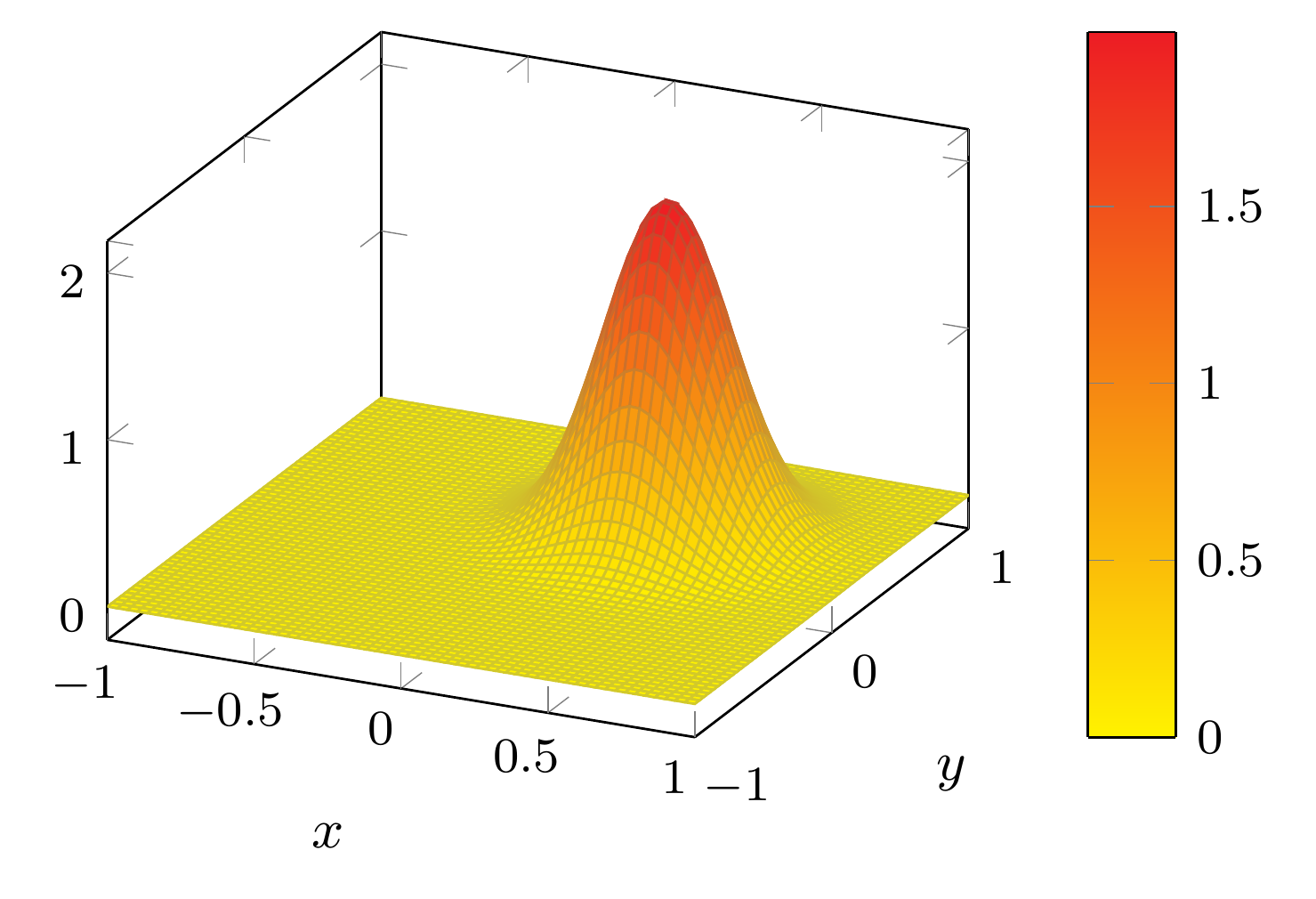}
		\caption{Intensity profile}
		\label{fig12}
	\end{subfigure}
	\caption{Fig.~\ref{fig11} The contours of incident light intensity and the resulting photodetections for a $4\times 4$ detector array. The red crosses represent the locations of signal photodetections during the observation interval, and the black ones correpond to noise. Fig.~\ref{fig12} shows the Gaussian profile of the intensity on the detector array.} \label{fig13}
\end{figure}

	Let us assume that the detector array, or the focal pixel array (FPA) as is also known in the literature, has $M$ square shaped detectors of uniform area.  The photodetection count $Z_m$ in the $m$th cell or detector of the array, resulting from the incident signal's energy, is modeled by a \emph{nonhomogeneous Poisson point process} during a given observation interval \cite{Bashir1}:
	\begin{align}
	P(\{ Z_m = z_m \}) = \frac{ \exp \left( {-\iint_{A_m} K \left[ \lambda_s(x,y, z) + \lambda_n \right] \, dx\, dy} \right) \left(\iint_{A_m} \left[ K \left(\lambda_s(x,y, z) + \lambda_n \right) \right] \, dx\, dy \right)^{z_m} } {z_m!} ,
	\end{align}
	where $A_m$ is the region of the $m$th detector on the FPA, and $Z_1, Z_2, \dots, Z_M$ are independent Poisson random variables. The function $\lambda_s(x, y)$ is the intensity profile of the incident optical signal, and $\lambda_n$ is a constant that represents the ``uniform'' intensity due to the background radiation and thermal effect of the array circuit. The quantity $z$ represents the distance between the transmitter and the receiver. The constant $K$ is required for the conversion of incident optical energy in Joules to the average number of photodetections resulting from that energy. It is given by$
	K = \frac{\eta \lambda T_p}{hc}$
	where $h$ is the \emph{Planck's constant} and its value is $6.62607004 \times 10^{-34}\, {m}^2 kg / s$, the constant $c$ is the speed of light in vacuum which is about $3\times 10^{8} \, m/s$, $\lambda$ is the wavelength of received light in meters, $\eta$ stands for the photoconversion efficiency, and $T_p$ represents signal pulse duration. The intensity profile for the signal over the detector array is assumed to be a Gaussian function in two dimensions \cite{Snyder}, \cite{Streit}:
	\begin{align}
	\lambda_s(x, y, z) \triangleq  I_0 \exp\left( \frac{-(x-x_0)^2 -(y-y_0)^2}{2 \rho^2(z)}  \right),
	\end{align}
	where $I_0$ is the peak intensity, $\rho(z)$ is the \emph{beam radius}, and $(x_0, y_0)$ is the center of the beam on the detector array. The factor $\rho(z)$ is defined as
	$\rho(z) \triangleq \rho_0 \sqrt{1 + \left(\frac{\lambda z}{\pi \rho_0^2} \right)^2}$, where $\rho_0$ is known as the \emph{beam waist}. Fig.~\ref{drone} illustrates a drone projecting a Gaussian beam on a detector array stationed on ground, and Fig.~\ref{fig13} outlines the Gaussian profile of the incident beam on the array of detectors.

	\section{Acquisition preliminaries} \label{acq_prelim}
	 Before the acquisition process can begin, the two terminals need to have some prior information about each others approximate location. Hence, depending on the accuracy of the prior information, and the errors associated with the pointing assemblies, there exists an uncertainty region in which a given terminal is expected to be present. In this paper, the term \emph{pointing} is used for initial pointing of the transmitter terminal towards the center of the uncertainty region.  
	 
	 {In the analysis that follows in this paper, we have assumed the receiver is stationary in space. Thus, the following analysis does not apply to receiver terminals that are mobile during the process of acquisition.}
	
	\subsection{Acquisition Uncertainty Region} \label{acq_uncertain}
	During the acquisition process, the transmitter has to scan a certain region in order to locate the receiver. The location of the receiver may be approximately known due to the localization algorithms that can position the receiver prior to the acquisition stage. Moreover, there may be errors in the pointing mechanisms of the transmitter---for instance, the gimbal pointing error or the gimbal jitter and calibration errors. All these errors lead to an \emph{uncertainty region} in which the receiver is known to be located. This uncertainty region is defined by the azimuth and elevation angle uncertainties. Fig.~\ref{spiral11} shows one example of the uncertainty region. The \emph{radial pointing error} is denoted by $\Phi$, and is defined to be the square root of  sum of squares of the \emph{elevation pointing error angle}, $\Phi_e$, and the \emph{azimuth pointing error angle}, $\Phi_a$ \cite{Kaushal}. Let us define $
	\Phi \triangleq \sqrt{\Phi_e^2 + \Phi_a^2}.
	$
	The angles $\Phi_e$ and $\Phi_a$ are assumed to follow a zero-mean Gaussian distribution\footnote{A Gaussian distribution is a good approximation for the error distribution (via the \emph{Lindeberg-Feller central limit theorem}) if we assume that the total error is a sum of a large number of independent random variables such as the gimbal calibration error, receiver localization error, error due to atmospheric turbulence, etc.}, and are modeled as independent random variables \cite{Bloom}:
	\begin{align} 
	p(\phi_e) = \frac{1}{\sqrt{2 \pi \sigma_e^2} } \exp\left( - \frac{(\phi_e)^2}{2 \sigma_e^2}   \right), \quad p(\phi_a) = \frac{1}{\sqrt{2 \pi \sigma_a^2} } \exp\left( - \frac{(\phi_a)^2}{2 \sigma_a^2}   \right),  \label{azi}
	\end{align}
	for $\phi_e > 0$ and $ \phi_a > 0$. Furthermore, we assume that $\sigma_e = \sigma_a.$ However, as shown in Section~\ref{GEDC}, any general case (elliptical uncertainty region) can be converted to the independent Gaussian error (circular uncertainty region) by an appropriate linear transformation, and the analysis of acquisition time concerning the circular uncertainty region will apply for any general uncertainty region.
	
	 The distribution of $\Phi$ is given by the well-known \emph{Rayleigh} distribution as $
	p(\phi) = \frac{\phi}{\sigma^2}\exp\left(- \frac{\phi^2 }{2 \sigma^2} \right) \label{radial} \text{ for }  0 < \phi < \infty,$
	where $\sigma = \sigma_e = \sigma_a$.

	\subsection{Acquisition Approach} \label{acq_approach}
	We will now discuss the initiation-acquisition protocol as discussed in \cite{Wang}. Let us assume that the two communicating parties have only a rough prior estimate of each other's position\footnote{A positioning system can provide rough location estimates.}.  Hence, the uncertainty region can be significantly large at this initial stage. During the first phase of this process, the initiator (terminal A) starts the acquisition process by scanning the uncertainty region (scan mode) with a beacon signal. The field-of-view or the beam radius may be set larger than normal during this initial scan in order to  locate the recipient (terminal B) as quickly as possible. Moreover, during this initial stage of acquisition, terminal A may also have to transmit a large amount of power in order to avoid being ``missed'' by the recipient\footnote{The recipient may not detect the initiator's signal if the transmitted power is not sufficient. }. The receiver of terminal B is in the ``stare'' mode since its ``listening'' for any transmitted signal. Once the beacon signal is detected by terminal B by virtue of a signal detection scheme, it sends a signal on an RF feedback channel to terminal A in order to halt the scanning process. Moreover, at this stage, it also acquires the bearings of terminal A by estimating the difference in its azimuth and elevation angles and those of terminal A\footnote{The process of acquiring the bearings is not discussed  in this paper but will be covered in an another future study on this topic.}. Once the difference in the angles is estimated by the receiver, the feedback control error signal activates the actuator circuits which leads to a motorized gimbal rotating the detector assembly in the appropriate direction in order to minimize the difference in angles. Thereafter, terminal B will start the same acquisition process by transmitting the beacon signal to terminal A. However, this time, the acquisition process will be much quicker since terminal B already has acquired the bearings of terminal A, and it can use a narrow beam to scan a smaller uncertainty region. Upon the detection of terminal B's signal by terminal A, the acquisition process is complete. Now the two terminals can start communicating data on the established optical link using the normal beam parameters (beam widths and power).

	\subsection{Scanning Techniques}
	The first step in locating the receiver is by steering the optical beam of the transmitter along a certain path (or pattern) in the uncertainty region. This process is known as \emph{scanning}. The following scanning techniques are typically used for the purpose of acquisition \cite{Kaushal}:
	\begin{enumerate*}
		\item \emph{Continuous spiral scan},
		\item \emph{step spiral scan},
		\item \emph{segmented scan},
		\item and \emph{raster scan}.
	\end{enumerate*} Continuous spiral is the most common and efficient scanning technique \cite{Kaushal}, \cite{XinLi}, \cite{Rzasa}. Therefore, in this paper, we will only discuss the continuous spiral scans. Having said that, the following discussion is general, and the techniques/results derived in the discussion can be applied to any scanning techniques with necessary modifications.
\begin{figure}
	\centering
	\begin{subfigure}[t]{2.5in}
		\centering
		\includegraphics[width=1.8in]{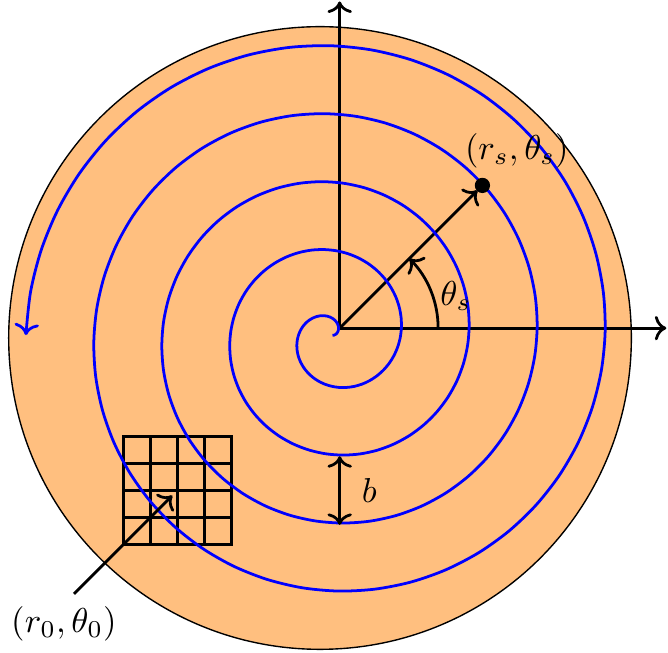} 
	\end{subfigure}
\vspace{-2cm} \hspace{2.5cm}
	\begin{subfigure}[t]{2.5in}
	\includegraphics[width=1.8in]{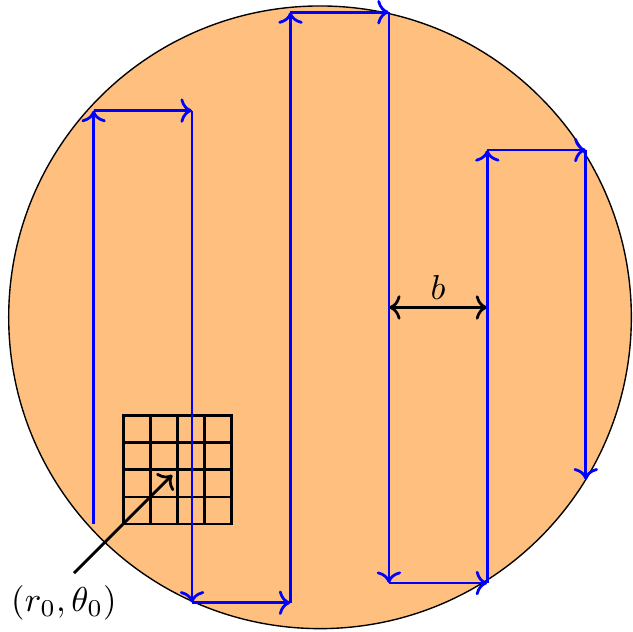}
	\end{subfigure}
\vspace{2.5cm}
	\caption{The spiral and raster scans for estimating the detector array's location. The uncertainty sphere is illustrated in orange.} \label{spiral11}
\end{figure}

In the continuous spiral scan, the transmitter begins the scan from the center of the uncertainty region, and move outwards in a spiral like fashion as shown in Fig.~\ref{spiral11}. This kind of spiral is known as the \emph{Archimedean spiral}.
The Archimedean spiral scan is the most efficient scanning technique in order to locate the receiver in the uncertainty region \cite{Kaushal}. This is true because of the circularly symmetric Gaussian distribution of $\Phi_e$ and $\Phi_a$. Hence, the closer a small differential region is to the center of the uncertainty region, the higher the probability that the receiver lies in that small region and vice versa.

The Archimedean spiral is defined in terms of polar coordinates as
$r_s = b \theta_s,$ where $b$ is the distance between the successive turnings in meters, $\theta_s$ is the angle in radians, and $r_s$ is the distance from the center of the spiral in meters. For the purpose of scanning, we divide the total length of the spiral into a number of uniform segments. In other words, we move on the spiral along discrete points. However, a uniform spacing between consecutive points on the spiral implies a nonuniform spacing for the discrete points of $\theta_s$\footnote{This fact can be observed in \eqref{spiral1} which says that when we move on the spiral in uniformly spaced steps away from the center of the uncertainty region, the radius increases which implies shrinkage in the step size of the angle $\theta_s$.}. If $\ell(t)$ represents any point on the spiral, then the distance between consecutive points on the ``discretized'' spiral is given by
\begin{align} \label{spiral1}
\ell[n] - \ell[n-1] {\approx} r_s[n] (\theta_s[n] - \theta_s[n-1]) , 
\end{align}
where $n$ stands for the discrete-time index and we assume that the step size ($\ell[n] - \ell[n-1]$) is small enough so that the left and right hand sides in \eqref{spiral1} are approximately equal. 

We can choose the step size, $\ell[n] - \ell[n-1]$,  and $b$  to be a suitable number that depends on the beam radius $\rho(z)$. This is important since we want the scanning beam footprints to have some overlap with each other as we move one step in time, and as we move from one turning to the next, on the spiral. This is consequential since we do not want to have any unscanned portions in the uncertainty region  after the scanning is complete.

The time required to go from the discrete-time index $n$ to $n+1$ (the time to move from one point on the spiral to the next) is known as the \emph{dwell time}. Let us denote the dwell time by $T_d$. The dwell time depends on the receiver processing (detection) time and the time-delay as given by $
T_d = T_r + \frac{ R}{c},$
where $T_r$ is the receiver processing time, $R$ is the (approximate) length of the optical link in meters, and $c$ represents the speed of light in space---$3 \times 10^8 \, m/s$. It should be noted that $T_r > T_{\text{obs}}$ where $T_{\text{obs}}$ is the observation interval for the detection problem.

Finally, the total number of steps in one scan of the uncertainty region can be upper bounded as
$
N_s = \left\lceil \frac{\text{area of uncertainty region}}{\text{area of the beam footprint}} \right\rceil  = \left\lceil \frac{ R_u^2}{ \rho^2(z)} \right\rceil,
$
where $R_u$ is the radius of the uncertainty sphere.

\subsection{General Error Distribution Case}\label{GEDC}
The analysis of the acquisition problem for a circular Gaussian error case is tractable to analyze since the distribution of the radius from the center of the uncertainty region is distributed as a Rayleigh random variable, and as we will see later in this paper, this translates to an exponential distribution (square of the Rayleigh random variable) for the acquisition time\footnote{Strictly, the distribution of the acquisition time only in the final scan is exponentially distributed}. The distribution of the acquisition time is hard to derive for a general Gaussian error distribution. However, if the covariance matrix of the error in $x$ and $y$ dimensions of the uncertainty region is known, then the analysis of acquisition time for the general (elliptical) Gaussian case  in two dimensions can be carried out in a circular Gaussian  domain by decorrelation/whitening approach. This is true because the distribution of acquisition time is the same in both domains if the transformation used has a unit determinant. This can be argued as follows. Let us assume the general case in which the uncertainty region is described by a Gaussian error distribution that has a general covariance matrix  $\mathbf{\Sigma}$, which is positive definite. For such a scenario, we can translate $\mathbf{\Sigma}$ to a diagonal covariance matrix with equal diagonal values by the linear transformation $\mathbf{T}$ of the data:
\begin{align}
\mathbf{T} \coloneqq \frac{\mathbf{\Lambda}^{-1/2} \mathbf{Q}^\intercal}{\det\left( \mathbf{\Lambda}^{-1/2} \mathbf{Q}^\intercal \right)},
\end{align}
where $\mathbf{Q}$ is the matrix containing eigenvectors of $\mathbf{\Sigma}$, and $\mathbf{\Lambda}$ is the (diagonal) eigenvalue matrix. It can be seen that the determinant of $\mathbf{T}$ is 1: thus, it would preserve volumes during transformation. 

\begin{figure}
	\centering
	\includegraphics[scale=0.6]{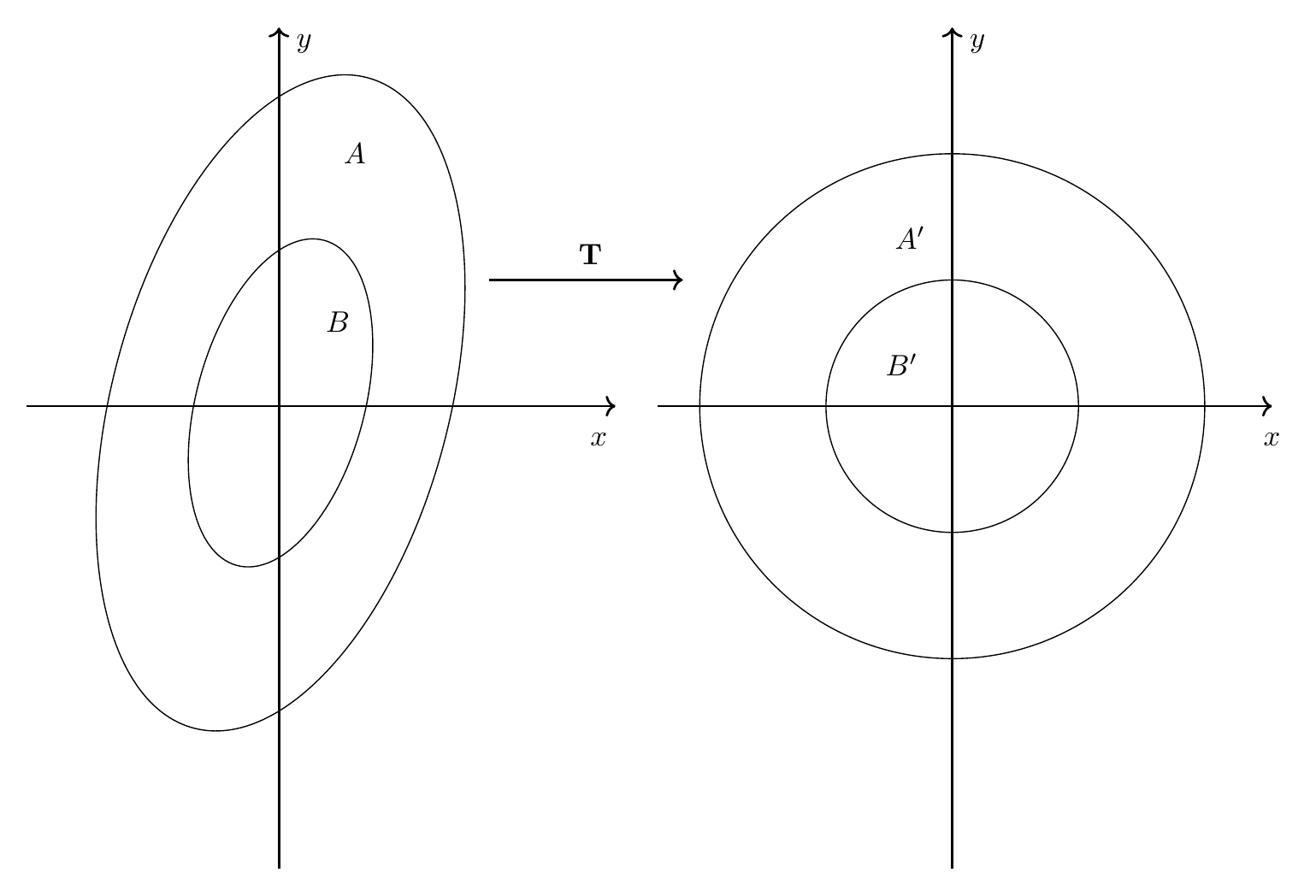}
	\caption{Transformation from elliptical to circular Gaussian}
	\label{fig111}
\end{figure}

{In this regard, without loss of generality, let us first assume that the probability of detection, $P_D$, is 1\footnote{$P_D=1$ represents the scenario when we are in a particular scan where we know that the receiver will be located ultimately in that scan.}. Additionally, let us assume that the general uncertainty region is given by ellipse $A$, and we want to find the probability that the acquisition time is less than some number $\tau_B$, where $\tau_B$ is the time required to scan the region $B \subset A$. The scanning is done in a spiral like fashion except that the paths traversed are elliptical contours (not circular as is the case with Archimedean spiral). Then, $\tau_B \approx T_d\times \left\lceil \frac{\text{vol}(B)}{\text{beam footprint}} \right\rceil$, and $
	P(\{ T \leq  \tau_B \}) =  {\iint_{B} f(x,y)\, dx \, dy}, \label{1}
	$
	where $f(x,y) \sim \mathcal{N}\left( \boldsymbol{0}, \mathbf{\Sigma} \right)$, $\iint_{B} f(x,y)\, dx \, dy$ is the probability that the receiver is located in region $B$, and $T_d$ is the dwell time. We ought to remember that vol$(B)  = \text{vol}(B') = \pi R_{B'}^2$ ($R_{B'}$ is the radius of $B'$, and beam footprint is $\pi \rho^2$) and $\text{vol}(A) = \text{vol}(A')$ because of the unit determinant of the transformation. Therefore, the time to scan region $B'$, $\tau_{B'} \approx T_d \times \left\lceil \frac{\text{vol}(B')}{\text{beam footprint}}\right\rceil \approx \tau_B$.
	Thus, for the circular Gaussian case, $
	P(\{ T' \leq  \tau_{B} \}) \approx P(\{ T' \leq  \tau_{B'} \}) =   {\iint_{B'} f'(x,y)\, dx \, dy},  \label{2}
	$
	where $f'(x,y)\sim \mathcal{N}\left(\boldsymbol{0}, \mathbf{\Sigma}' \right),$ $\sqrt{\mathbf{\Sigma}'} = \mathbf{T} \sqrt{\mathbf{\Sigma}}$, and $\mathbf{\Sigma}'$ is diagonal with equal values on the diagonal. The quantity $\sqrt\mathbf{\Sigma}$ is the element-wise square root of matrix $\mathbf{\Sigma}$.
	The two distributions---$P(\{ T \leq  \tau_{B} \})$ and $P(\{ T' \leq  \tau_{B} \})$---will be equal if we can show that, 
	\begin{align}
	\iint_{B} f(x,y)\, dx \, dy = \iint_{B'} f'(x,y)\, dx \, dy. \label{3}
	\end{align}
	It is not straightforward to prove \eqref{3} rigorously. However, with simulations,  the equality  in \eqref{3} can be verified for any general Gaussian function $f$ and any volume $B$ (which lead to their corresponding $f'$ and $B'$). Thus, the distribution of the acquisition time does not change if we translate the problem from the general uncertainty region to a circular uncertainty region.}

Finally, in order to make the argument more precise, we  have to address the issue of the beam profile after transformation. A circular (Gaussian) beam profile is used to scan the elliptical uncertainty region. After transformation, the uncertainty region is transformed, but the circular profile of the beam is retained because the beam profile can't be changed at the transmitter. This may lead to the fact that the  time to scan $B$, $\tau_B$, and the time to scan $B'$, $\tau_{B'}$, may be slightly different because the packing efficiency of small circles (beam footprint) packing the circle $B'$  is not the same as the circles packing an ellipse $B$. In order to fix this issue, let us define  $\tau^{(B)} \coloneqq a T_d \left\lceil ({R_{B'}^2}{/\rho^2}) \right\rceil$, where $a$ is an appropriate scaling factor greater than 1. Now, for instance, if we choose $a=2$, then it is easy to see that $\tau_B, \tau_{B'} < \tau^{(B)}$. In other words, $\tau^{(B)}$ is an upper bound on the time to scan either region $B$, or $B'$, and we can use this upper bound for the optimization of the acquisition time as a function of beam radius and the number of detectors $M$ in the array. The scaling of time by a factor of $a$ is not going to affect the fundamental results of this analysis. For the sake of simplicity, we have set $a=1$ in this paper.

%Alternatively, we can pack twice as many beam footprints as needed in $B$ or $B'$, i.e., $a T_d \left\lceil \frac{R_B^2}{\rho^2}\right\rceil$, with $a=2$. This can be done by shortening the step size $\ell[n]-\ell[n-1]$ on the spiral, or reducing $b$, or both. This will guarantee that the time $\tau_B$ and $\tau_{B'}$ are equalized to $\tau^{(B)}$. However, due to the possible overlaps of beam footprints during scanning, the probability of detection will be greater than the one-shot probability of detection. Therefore, using the one-shot probability of detection in the analysis will  increase the mean acquisition time. However, as discussed earlier, we are analyzing the optimization of an upper bound  on acquisition time with respect to beam radius and $M$, and the optimization results will not be affected by loosening the upper bound on acquisition time. 

\section{Acquisition Process} \label{acq_process}
In this section, we review the detection scheme required for the acquisition problem. In this regard, we assume that the transmitter and receiver have already been synchronized in time through a low bandwidth radio frequency channel. Such a channel is also used for exchange of control information between the transmitter and the receiver.

The technique that we use for acquisition is as follows. At a given point on the scanning spiral, the transmitter sends the beacon signal which is a pulse of duration $T_p$.  We set the observation interval $T_{\text{obs}} > T_p+R/c$ in order to account for the time-delay. The intensity of the pulse has a Gaussian profile in space as discussed in Section~\ref{profile}. The receiver makes use of a detection algorithm in order to detect the presence or absence of the transmitted signal. If no signal is detected during the entire scan of the uncertainty region, the scanning process is reinitiated from the center of the uncertainty region, and this process repeats until the receiver detects the transmitted signal. The point on the scanning path at which the receiver detects the signal is then conveyed back to the transmitter on the control channel so that the transmitter can now ``point'' in the right direction.

 In the following section, we explain the detection algorithm for the acquisition stage, and how this acquisition stage can be optimized. 

\subsection{Detection Algorithm for Acquisition}\label{mldetector}
When the footprint of the beam partially or completely covers the detector array, we observe signal photons as well as noise photons during the observation interval. In order to detect the presence of the optical signal on the array, we carry out a likelihood ratio test that decides between the hypothesis
	$H_0$: the signal beam is absent on the array, 
versus the hypothesis\newline
	$H_1$: there is a signal beam present on the array.

If we have perfect knowledge of the beam parameters and the orientations of the transmitter and the receiver, then a maximum likelihood detector can be defined as 
\begin{align}
&\sum_{m=1}^{M}   z_m \left[ \ln  \left( \iint_{A_m}  K [\lambda_s(x,y, z) + \lambda_n]  \, dx\, dy\right) - \ln \left( K \lambda_n |A_m|  \right)  \right]  \underset{H_1}{\overset{H_0}{\lessgtr} } \gamma +  \left( {\iint_{\mathcal{A}} K  \lambda_s(x,y, z)  \, dx\, dy} \right), \nonumber\\
& \implies \sum_{m=1}^{M}   z_m \left[ \ln  \left( 2\pi \rho^2 K I_0 \left(\Phi(u_{mx}) - \Phi(l_{mx}) \right) \left(\Phi(u_{my}) - \Phi(l_{my}) \right) + K \lambda_n |A_m|\right) - \ln \left( K \lambda_n |A_m|  \right)  \right] \nonumber \\
& \underset{H_1}{\overset{H_0}{\lessgtr} } \gamma +  2\pi \rho^2 K I_0 ,  \label{detect}
\end{align}
where $\mathcal{A}$ is the region of the detector array, that is, $ \mathcal{A} \triangleq \bigcup_{m=1}^M A_m$, and $\gamma$ is a threshold that is set according to some desired probability of false alarm. The quantity $\Phi$ represents the cumulative distribution function of a standard normal random variable,  and $(u_{mx}, u_{my}), (u_{mx}, l_{my}), (l_{mx}, u_{my})$ and $(l_{mx}, l_{my})$ represent the coordinates of the four corners of the detector region $A_m$. The function $\lambda_s(\cdot)$ and the constants $\lambda_n$ and $K$ are defined in Section~\ref{profile}.
\subsection{Probabilities of Missed Detection and False Alarm}
For the sake of clarity, let us denote the factors \newline $\ln  \left( \iint_{A_m}  K [\lambda_s(x,y, z) + \lambda_n]  \, dx\, dy\right) - \ln \left( K \lambda_n |A_m|  \right)$ and $\gamma +  \left( {\iint_{\mathcal{A}} K  \lambda_s(x,y, z)  \, dx\, dy} \right)$ in \eqref{detect} by $\alpha_m$ and $\gamma_0$, respectively. The factor $\alpha_m$ captures the SNR content in the $m$th detector of the array if we rewrite its expression as $
\alpha_m =  \ln (1 + \text{SNR}_m), $
where $\text{SNR}_m \triangleq \frac{\iint_{A_m}  \lambda_s(x, y, z) \, dx \, dy} {\lambda_n |A_m|}$.
Then, the one-shot probability of missed detection for the observation interval $T_p$ (or for a fixed point on the scanning path), denoted by $P_{m}$, can be written as
$
P_m \triangleq P\left(  \left\{   \sum_{m=1}^M Z_m \alpha_m  < \gamma_0    \right\}  \right), \label{Pm}
$
where $Z_m \sim \text{Poisson}\left(\iint_{A_m} K(\lambda_s(x,y, z)+ \lambda_n) \, dx \, dy \right)$. The random variable $Y_s \triangleq \sum_{m=1}^MZ_m \alpha_m$ is a linear combination of independent Poisson random variables whose density function is not straightforward to compute exactly. However, we know the mean and variance of $Y_s$ as $
\mu_s \triangleq \sum_{m=1}^M  \alpha_m K \iint_{A_m} (\lambda_s(x,y, z)+ \lambda_n) \, dx \, dy,$
 and variance $
 \sigma^2_{s}\triangleq \sum_{m=1}^M  \alpha_m^2 K \iint_{A_m} (\lambda_s(x,y, z)+ \lambda_n) \, dx \, dy.$
 
 We will now look at two approximations of $P_m$ in the following sections. 
 
 \subsubsection{Gaussian Approximation}
 For the special case of  large number of detectors in the array, and for a large $\lambda_n$ (or/and a large $I_0$ and $\rho(z)$), we can approximate the random variable $Y_s$ by a Gaussian random variable via the \emph{Lindeberg-Feller central limit theorem}. Then, $P_m$ can be approximated asymptotically as $
P_m \approx 1- Q\left(  \frac{\gamma_0 - \mu_{s} } {\sigma_{s}}   \right),$
where the $Q(x)$ is defined as the probability that a standard normal random variable is greater than a real number $x$.

\subsubsection{Poisson Approximation}
Another approximation is via the scaled Poisson random  variable (scaled by the factor $k_s$) as follows \cite{Tan}:
\begin{align}
P_m = P\left(  \left\{   \sum_{m=1}^M Z_m \alpha_m  < \gamma_0    \right\}  \right) \approx P\left(  \left\{ Z_0 < k_s \gamma_0 \right\} \right) = Q(\lfloor k_s\gamma_0 + 1 \rfloor, k_s \mu_s),
\end{align}
where $k_s \triangleq \frac{\mu_s}{\sigma^2_s}$, and $Z_0$ is Poisson$(k_s\mu_s)$. The function $Q(x, y)$ is known as the \emph{regularized Gamma function} and is defined as 
$
Q(x, y) \triangleq \frac{\Gamma(x,y)}{\Gamma(x)}, \label{rgamma}
$
where $\Gamma(x, y)$ is the \emph{upper incomplete Gamma function}:
$
\Gamma(x, y) \triangleq   \int_{y}^{\infty} t^{x-1} e^{-t}\, dt,
$
and $ \Gamma(x) \triangleq \int_{0}^{\infty} t^{x-1} e^{-t}\, dt$.

\begin{figure}
	\centering
	\includegraphics[scale=0.8]{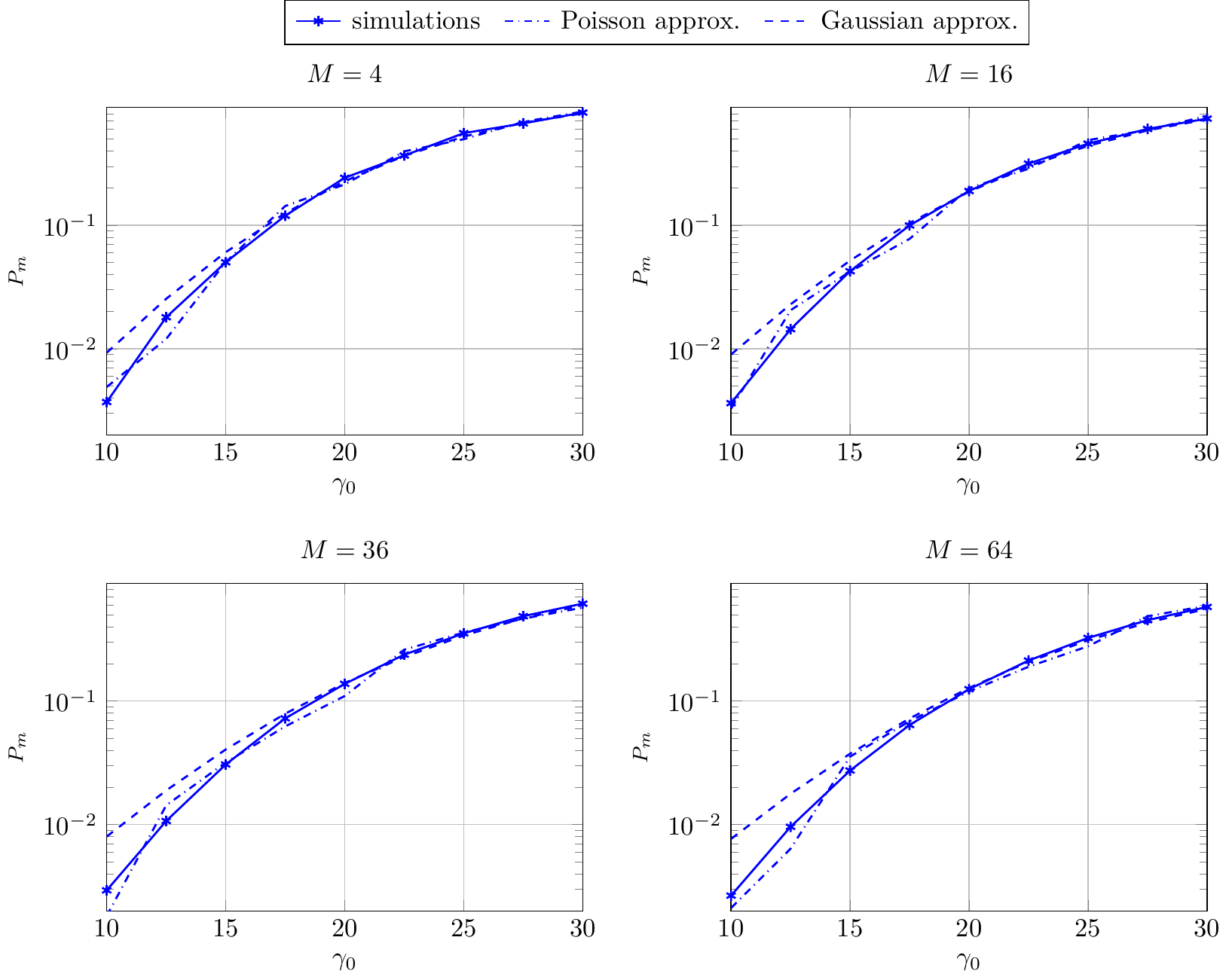}
	\caption{Difference between the true value of $P_m$ and its approximations for the received signal power of $3.25 \times 10^{-7} $ W, noise power of $3 \times 10^{-7}$ W, $\rho(z) = 0.2$ m, $|C| = 4$ $\text{m}^2$, and $(x_0, y_0)=(0.4, 0.4)$.} \label{approx1}
\end{figure}
Fig.~\ref{approx1} and Fig.~\ref{approx2} depict the estimate of $P_m$ and the two approximations. The true value of $P_m$ is estimated by using Monte Carlo experiments. It can be seen that the scaled Poisson distribution provides a better tail probability approximation than the Gaussian case. 

\subsubsection{Bounds for the Probability of Missed Detection}
The probability of missed detection for one scan, denoted by $P_M$, is defined as the probability of the event that the transmitter fails to locate the receiver in one full scan of the uncertainty region when the receiver is known to be present in the said uncertainty region. Before we present the bounds on $P_M$, we like to define the integers $N_0$ and $N_1$ via the \emph{circle packing a square} argument in the following analysis. For \emph{square packing}, the fraction of the area of a square covered by any collection of circles of uniform radius is $\pi /4$. For \emph{hexagonal packing}, the fraction of the area covered is $\pi \sqrt{3}/6$. The number $N_0$ and $N_1$ are positive integers such that 
\begin{figure}
	\hspace{1cm}\begin{minipage}{0.5\linewidth}
		\centering
		\includegraphics[scale=0.5]{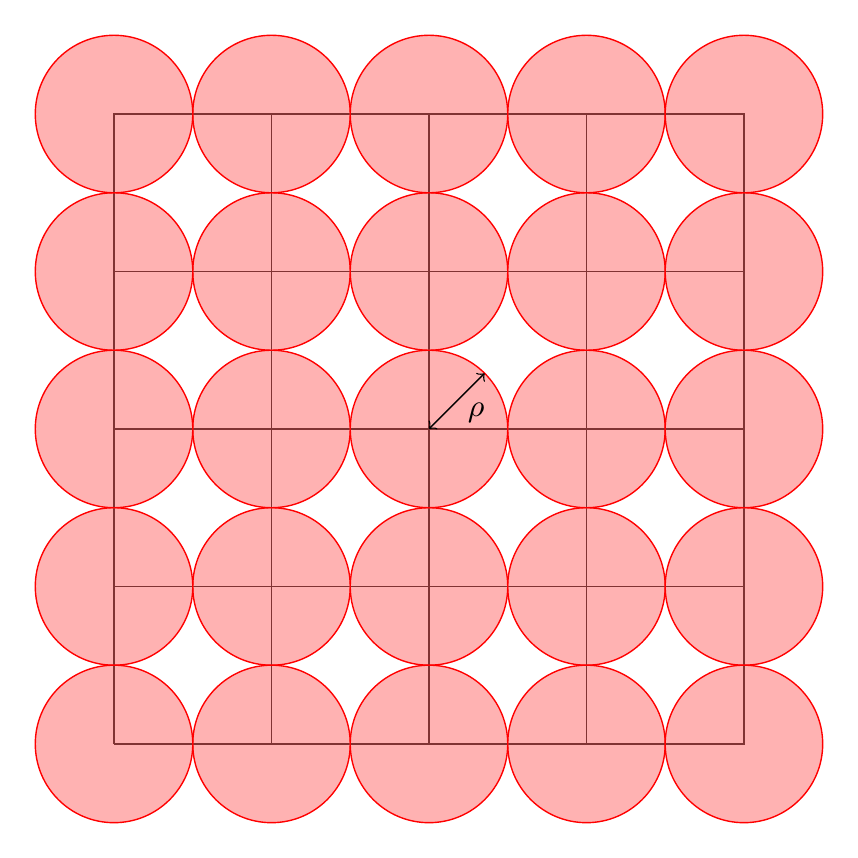}
	\end{minipage}
%\hspace{6cm} \vspace{-1.5cm}
\begin{minipage}{0.5\linewidth}
	\centering
	\includegraphics[scale=0.5]{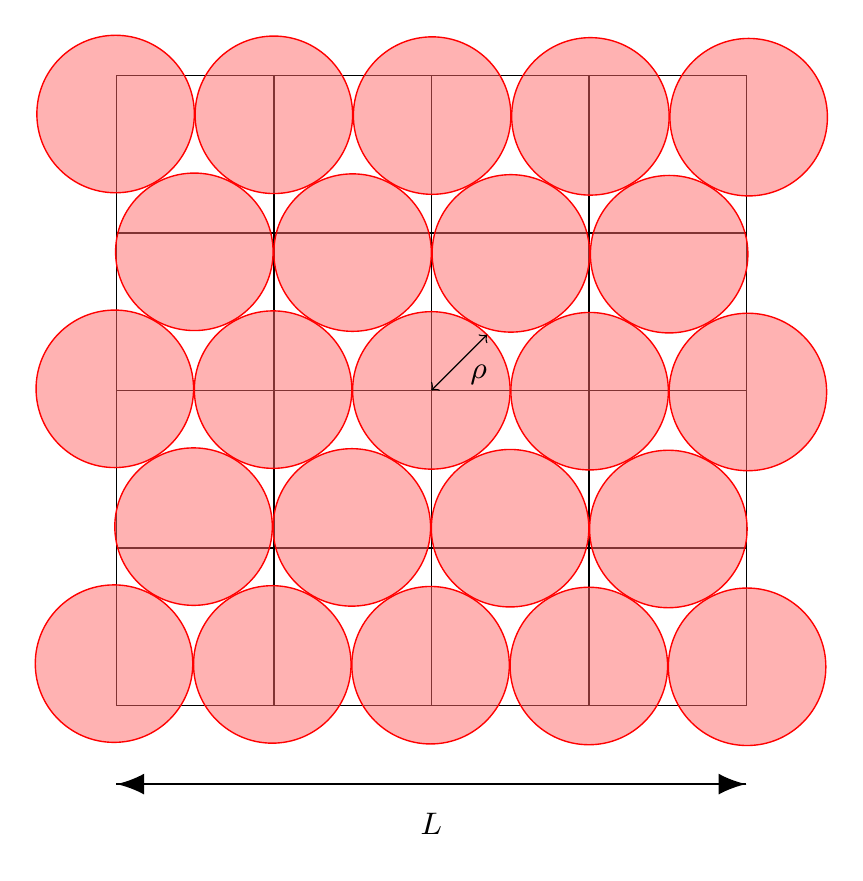}
\end{minipage}
 \caption{Square packing (left) versus hexagonal packing (right). The hexagonal packing utilizes the space most efficiently.} \label{circ_packing}

\end{figure}
$
N_0 \triangleq \left\lfloor \frac{|\mathcal{A}| \pi/4}  {\pi \rho^2(z)}\right\rfloor = \left\lfloor \frac{|\mathcal{A}|}{4 \rho^2(z)} \right\rfloor, \text{ and }
N_1 \triangleq \left\lceil \frac{|\mathcal{A}| \pi \sqrt{3}/6}{\pi \rho^2(z)} \right\rceil= \left\lceil \frac{|\mathcal{A}| \sqrt{3}}{6 \rho^2(z)} \right\rceil.
$
In other words, the number of circles that can be packed  in a square of area $|C|$ for square packing is either $N_0$ or $N_0 + 1$, and the number of circles packed in the similar sized square for hexagonal packing is either $N_1$ or $N_1 - 1$ (see Fig.~\ref{circ_packing}). The number $\lfloor N_0 - L/\rho(z)\rfloor$ in \eqref{PM} is the lower bound on the number of circles that will have a full overlap with the detector array at any point on the scanning path. It is also a fact that $N_0 < N_1$.

 The probability $P_M$ is bounded as
\begin{align}
P_{M_L} < P_M < P_{M_U}, \label{PM}
\end{align}
where $P_{M_L} \triangleq P_m^{N_1}$,  $P_{M_U} \triangleq P_m^{\lfloor N_0 - L/ \rho(z)\rfloor}$, $L$ is the length of one side of the square-shaped detector array, and $\frac{L}{\rho(z)}$ is the maximum number of instances when the beam footprint has a partial overlap with the array.  Moreover,  for the bound in \eqref{PM}, the inherent assumption is that any two or more events related to missed detection are independent of each other.

\begin{figure}
	\centering
	\includegraphics[scale=0.8]{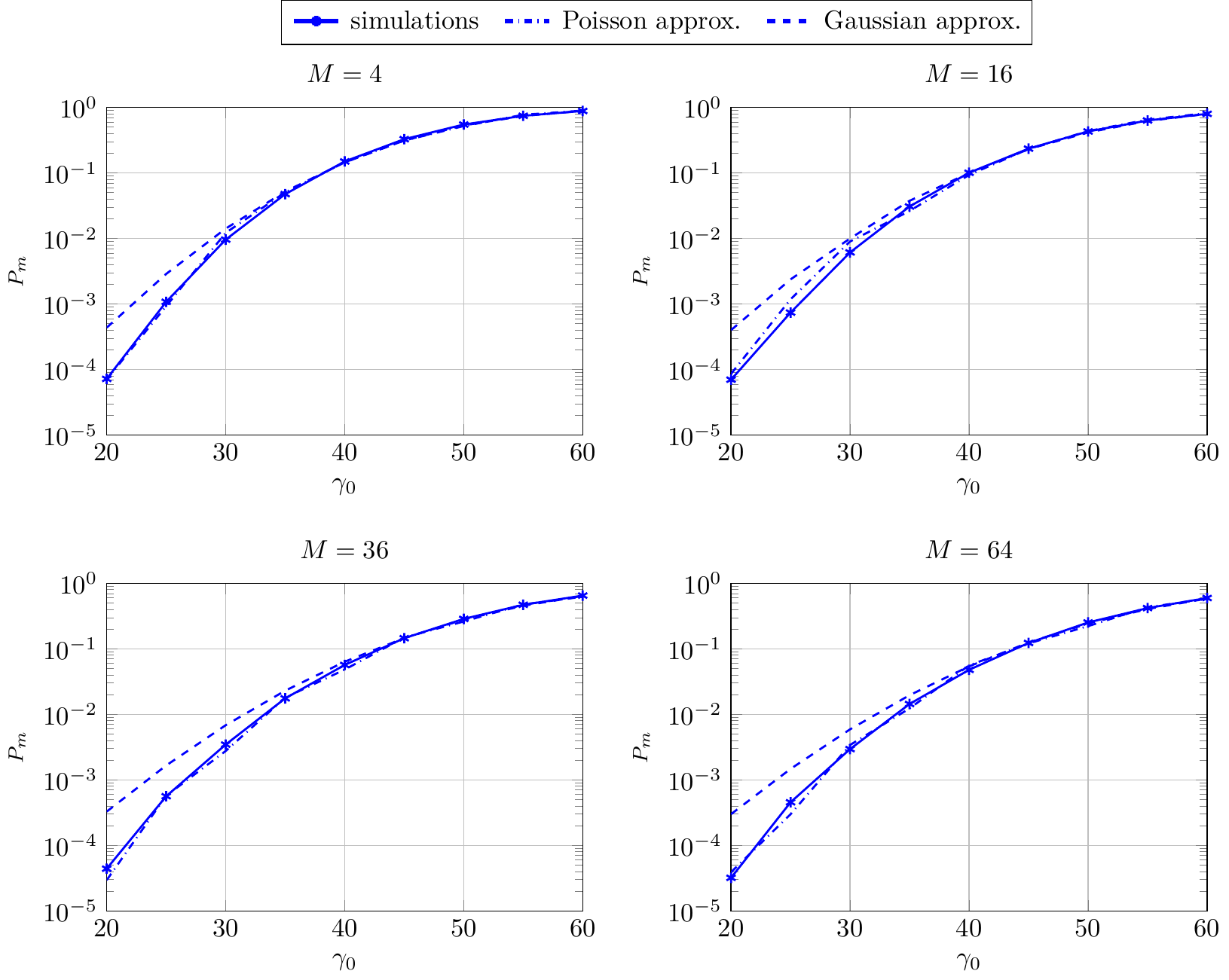}
	\vspace{-0.2cm}
	\caption{Difference between the true value of $P_m$ and its approximations for the received signal power of $6.5 \times 10^{-7} $ W, noise power of $6 \times 10^{-7}$ W, $\rho(z) = 0.2$ m, $|C| = 4$ $\text{m}^2$, and $(x_0, y_0)=(0.4, 0.4)$.} \label{approx2}
\end{figure}

\subsubsection{Bounds for the Probability of False Alarm}
The (one-shot) probability of false alarm at one point of the scanning path, denoted by $P_{f}$, is
\begin{align}
P_{f} = P\left(  \left\{   \sum_{m=1}^M Z_m \alpha_m  > \gamma_0    \right\}  \right) \approx 1 - Q(\lfloor k_n \gamma_0 + 1 \rfloor, k_n \mu_n) \label{PF}
\end{align}
for the case when $Z_m \sim \text{Poisson}(K\lambda_n |A_m|)$. The regularized gamma function $Q(x, y)$ is defined in \eqref{rgamma}, and
$\displaystyle \mu_n \triangleq  \sum_{m=1}^M \alpha_m K \lambda_n |A_m|,
\sigma^2_n \triangleq  \sum_{m=1}^M \alpha_m^2 K \lambda_n |A_m|, $ and $
k_n  \triangleq \frac{\sum_{m=1}^{M}\alpha_m}{\sum_{m=1}^{M}\alpha_m^2}.$ {In order to set the threshold $\gamma_0$, we fix the left hand side of \eqref{PF} (which is $P_f$) to a certain (small) value that is required by the system. Thereafter, the value of $\gamma_0$, that makes the two sides of \eqref{PF} equal to each other, is the value of the chosen threshold.}

\begin{figure}
	\centering
	\begin{subfigure}[b]{0.35\textwidth}
		\includegraphics[width=\textwidth]{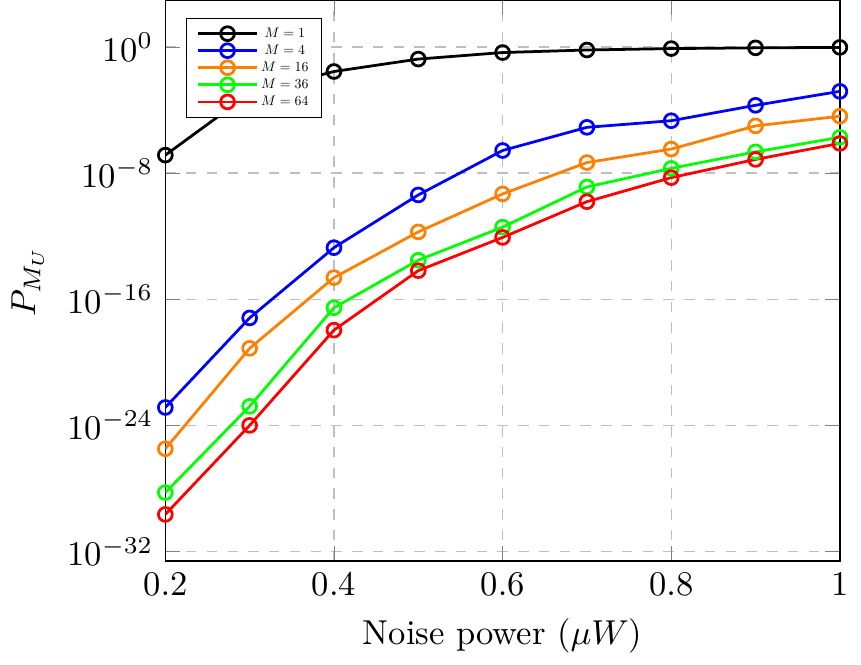}
		\caption{$\rho(z)=0.25$ m}
		\label{fig1}
	\end{subfigure}
	\begin{subfigure}[b]{0.35\textwidth}
		\includegraphics[width=\textwidth]{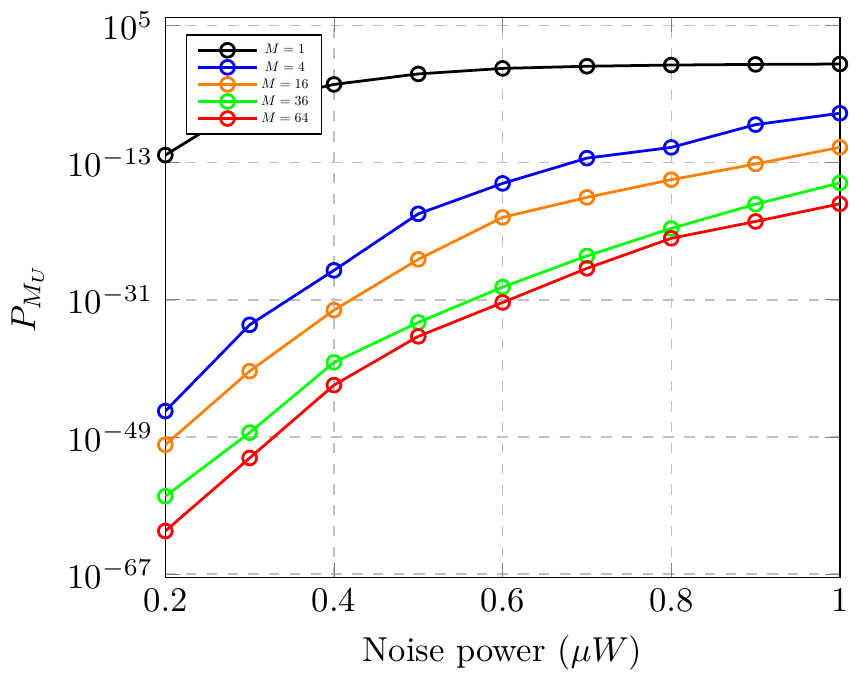}
		\caption{$\rho(z)=0.2$ m}
	\label{fig2}
	\end{subfigure}\vspace{-0.3cm}
\caption{The upper bound on the probability of missed detection as a function of noise power for different detector arrays. The signal power is 1 $\mu W$ and the beam radius $\rho(z)=0.25$ m in Fig.~\ref{fig1}. The threshold is chosen so that the upper bound on probability of false alarm is $4.44 \times 10^{-12}$.  For Fig.~\ref{fig2}, the signal power is 1 $\mu W$ and the beam radius $\rho(z)=0.2$ m. The threshold is chosen so that the upper bound on probability of false alarm is $6.937 \times 10^{-12}$. } \label{PM1}
\end{figure}

\begin{figure}
	\centering
	\begin{subfigure}[b]{0.35\textwidth}
		\includegraphics[width=\textwidth]{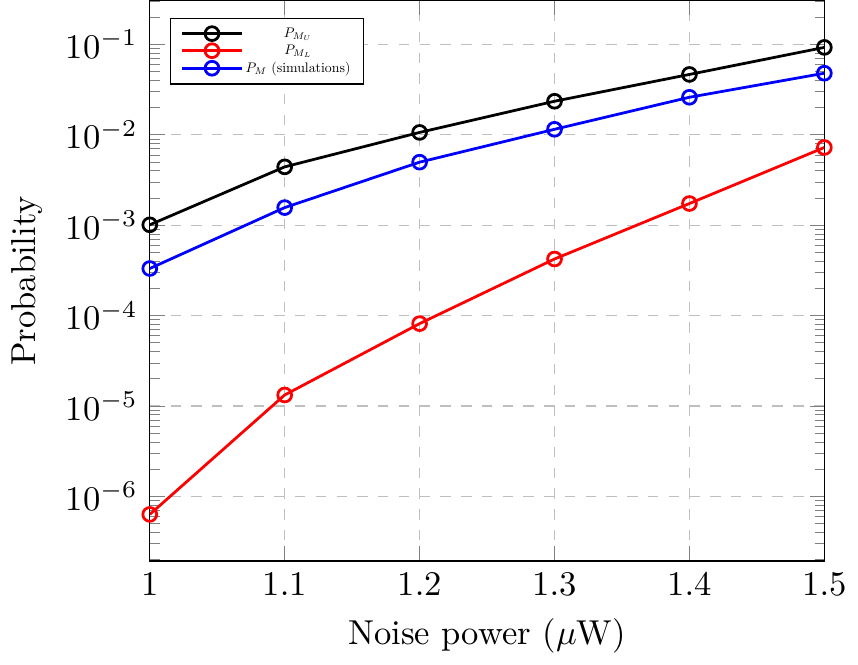}
		\caption{$M=16$}
		\label{fig5}
	\end{subfigure}
	\begin{subfigure}[b]{0.35\textwidth}
		\includegraphics[width=\textwidth]{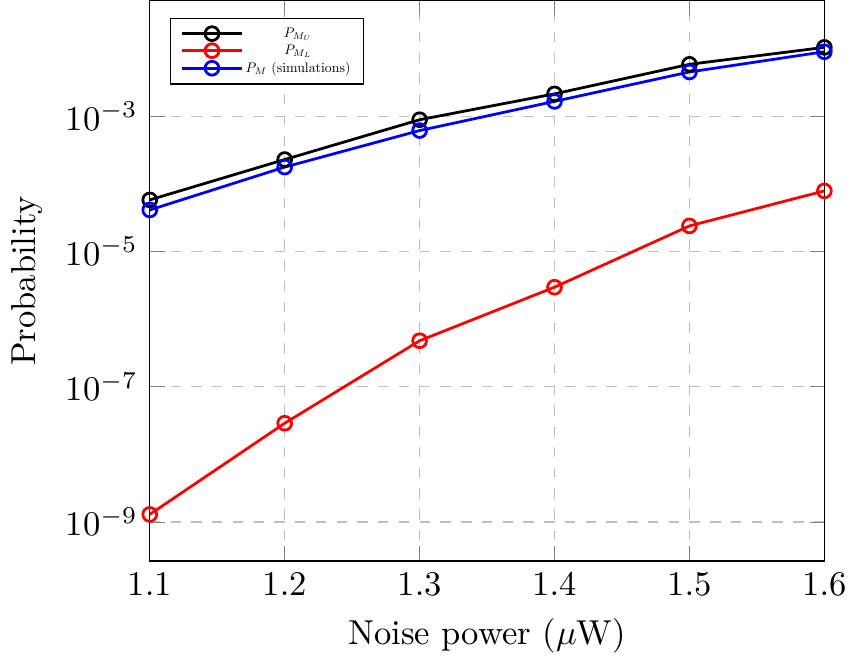}
		\caption{$M=36$}
		\label{fig6}
	\end{subfigure} \vspace{-0.3cm}
	\caption{The upper and lower bounds on the probability of missed detection versus the probability of missed detection estimated through Monte Carlo simulations for two types of detector arrays. The signal power is 0.63 $\mu W$ and the beam radius $\rho(z)=0.2$ m for both the figures. } \label{PM2}
\end{figure}

Finally, the probability of false alarm for one complete scan (we denote this by $P_F$), given that the array is not detected during such scan, is bounded as follows.
\begin{align}
P_{F_L} < P_F < P_{F_U},
\end{align}
where $P_{F_L} \triangleq 1-(1-P_{f})^{N_s - N_1}  $ and $P_{F_U} \triangleq 1- (1-P_f)^{ N_s - \left\lfloor N_0 - \frac{ L} {\rho(z)}  \right\rfloor  }$. A simple proof in this regard is as follows:
\begin{align}
P(\text{No false alarm in one scan}) = 1 - P_F > (1-P_f)^{N_s - \left\lfloor N_0 - \frac{L}{\rho(z)}\right\rfloor}
\end{align}
and 
\begin{align}
&1-P_F < (1 - P_{f})^{N_s - N_1} 
 \implies 1 - (1 - P_{f})^{N_s - N_1} < P_F < 1 - (1-P_f)^{N_s - \left\lfloor N_0 - \frac{L}{\rho(z)}\right\rfloor}.
\end{align}
Fig.~\ref{PM1} shows the graphs of $P_{M_U}$ for different types of detector arrays. Moreover, Fig.~\ref{PM2} compares the upper and lower bounds ($P_{M_U}$ and $P_{M_L}$) with the value of $P_M$ estimated with simulations for $M=16$ and $M=36$ array. It can be noticed that the estimated value of $P_M$ is closer to the upper bound $P_{M_U}$ for the two types of detector arrays. 

\subsection{Maximum Likelihood Detector Performance for Uncertain Beam Parameters}
In deriving the performance measures of the maximum likelihood detector in Section~\ref{mldetector}, we assumed perfect knowledge of beam parameters such as $I_0$, $\lambda_n$ and $\rho(z)$, %and the difference in azimuth and elevation angles, $|\Delta \theta_e|$ and $\Delta \theta_a$ were assumed to be zero.
 However, turbulent and scattering channels can cause random beam spreading and signal attenuation. %Moreover, if the localization system for initial pointing is accurate enough, then $|\Delta \theta_e|$ and $|\Delta \theta_a|$ are approximately zero. However, in a general setting, we have to account for nonzero $|\Delta \theta_e|$ and $|\Delta \theta_a|$ as well.
  Therefore, in this section, we use simulations in order to ascertain the maximum likelihood detector's performance in terms of $P_m$ when there is uncertainty in beam parameters.
  
\begin{figure}
	\centering
	\includegraphics[scale=0.75]{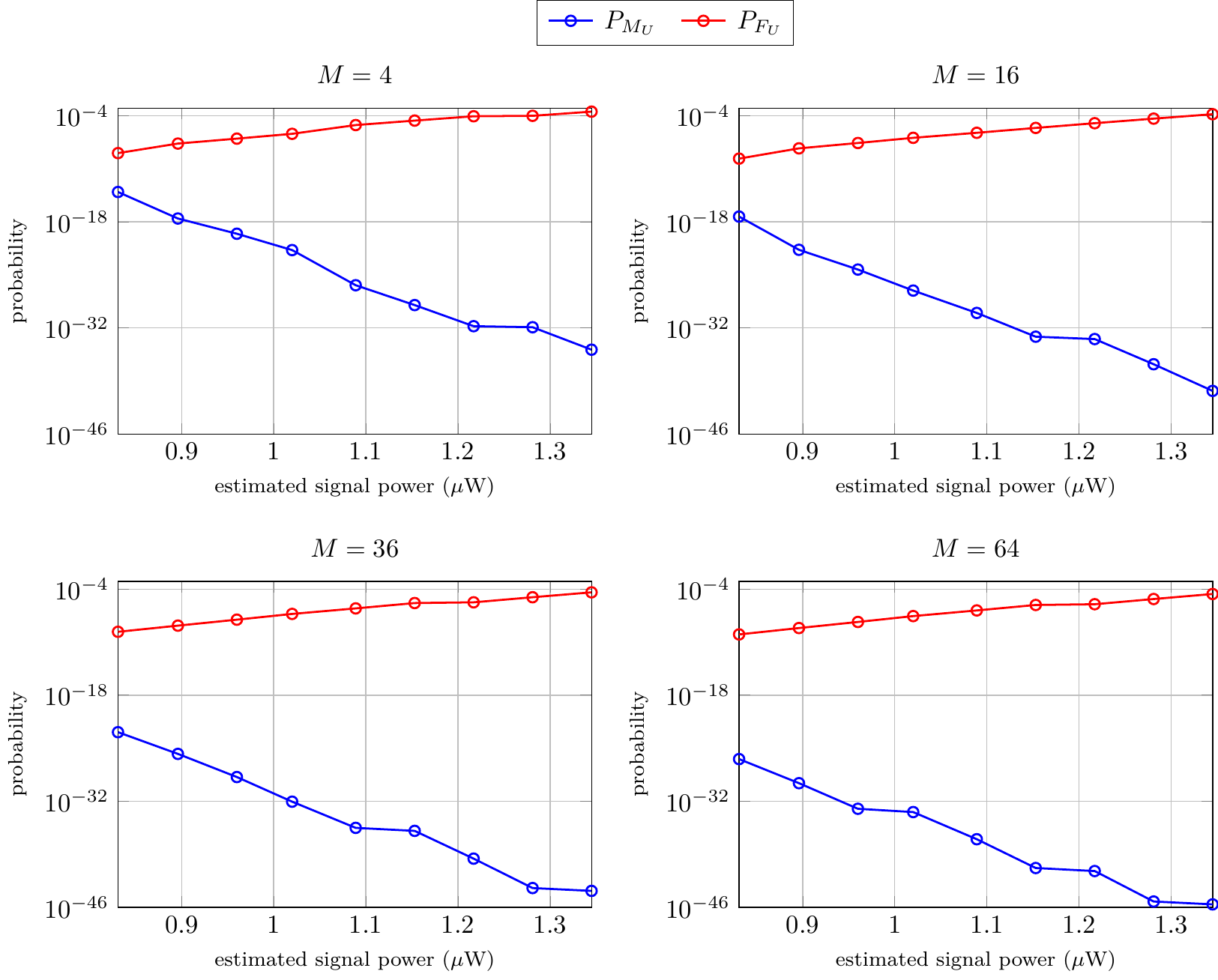}
	\caption{The upper bounds on the probability of missed detection and probability of false alarm as a function of estimated signal power for four different detector arrays. The true received signal power was $1.089$ $\mu$W, and the noise power was 1 $\mu$W. The threshold was set at $60$.} \label{I_0_un}
\end{figure}

\begin{figure}
	\centering
	\includegraphics[scale=0.75]{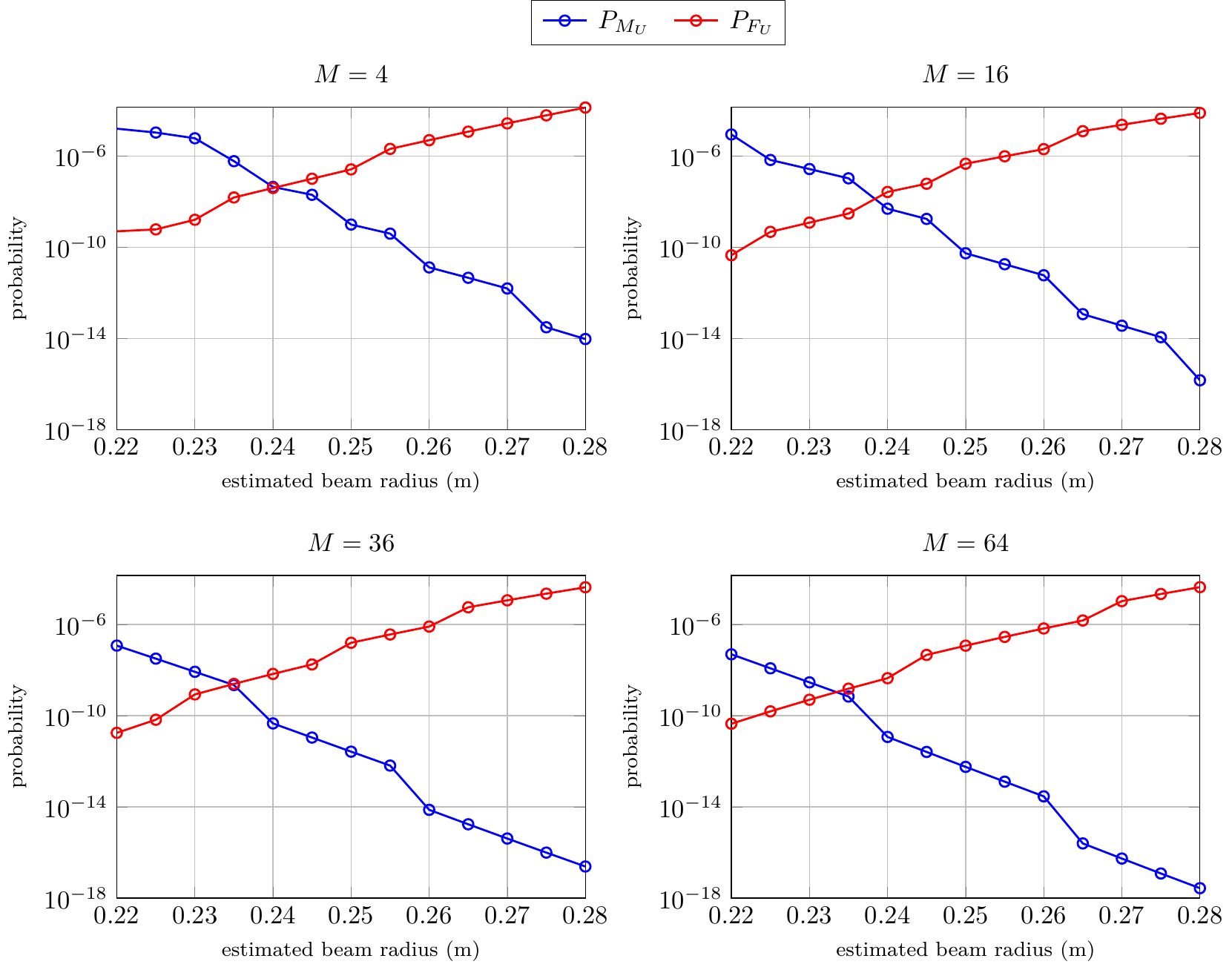}
	\caption{The upper bounds on the probability of missed detection and probability of false alarm as a function of estimated beam raduis $\rho(z)$ for four different detector arrays. The true beam radius was 0.25 m, and the noise power was 1 $\mu$W. The threshold was set at $60$.} \label{p_un}
\end{figure}

  As can be observed in Fig.~\ref{I_0_un} and Fig.~\ref{p_un}, the general effect of overestimating $I_0$ and $\rho(z)$ results in an increase in the probability of false alarm and a reduction in the probability of missed detection. This is explained by an increase in the value of factors $\alpha_m$ which cause an increase in false alarm probability, and a decline in the missed detection probability. However, this behavior is different in case of the beam center position estimation, in which the probability of missed detection is minimized only at the true value of beam position. The effect of beam position estimation is negligible in case of false alarm probability since the noise photons do not ``favor'' one beam position value over another for the evaluation of sufficient statistic. However, the dip in the false alarm probability near the edge of the array is observed because of a diminishing of the values of some of $\alpha_m$'s.
  
\begin{figure}
	\centering
	\includegraphics[scale=0.75]{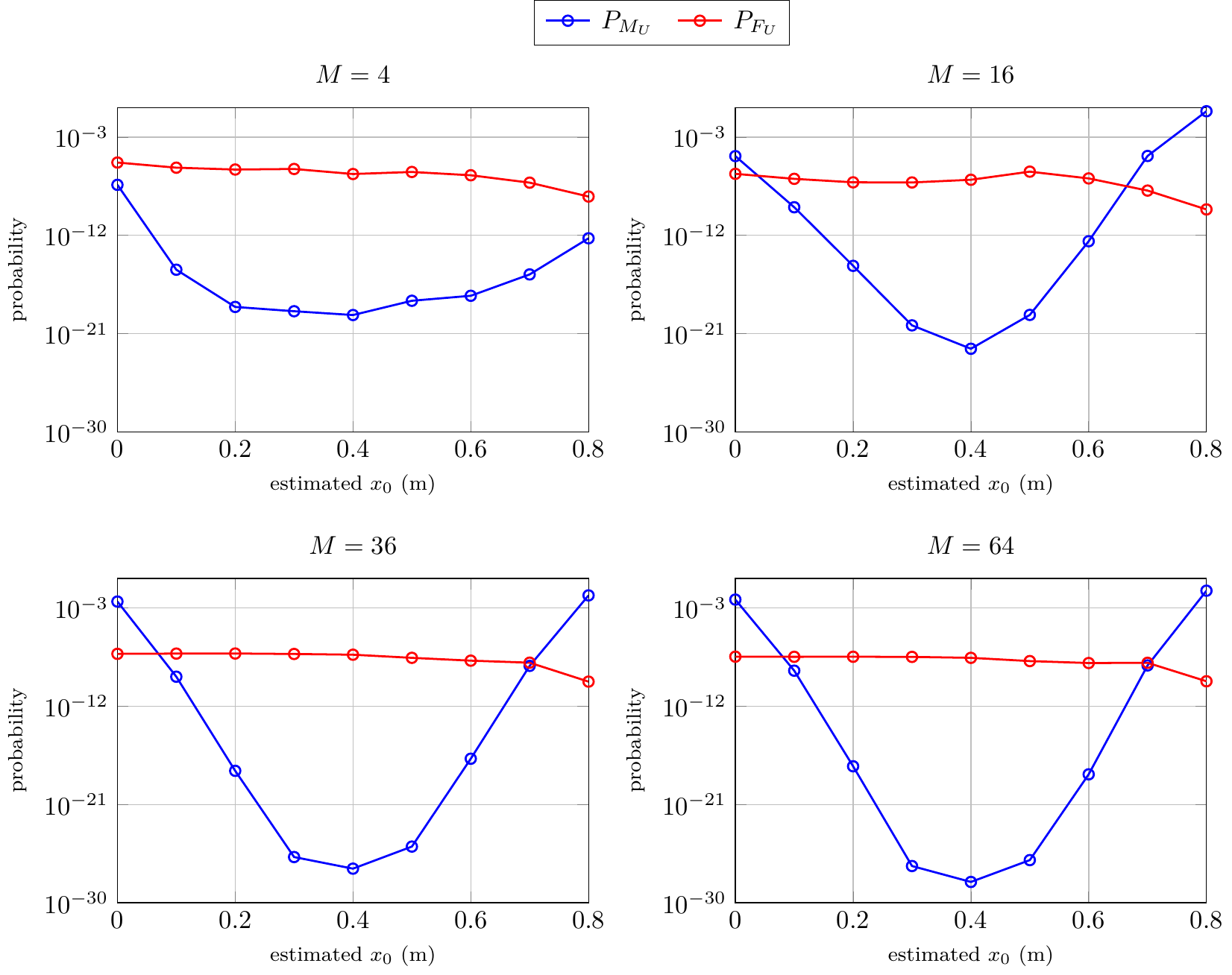}
	\caption{The upper bounds on the probabilities of missed detection and false alarm as a function of estimated $x_0$ for four different detector arrays. The true value of $x_0$ was $0.4$, the beam radius was $0.2$ m, and the noise power was $1$ $\mu$W. The threshold was set at 60, and the received signal power was $1$ $\mu$W.} \label{center_un}
\end{figure}

  {Fig.~\ref{center_un} depicts the ``one dimensional'' effect of the estimated beam center on the detector array on the probabilities of missed detection and false alarm for different types of detector arrays. As can be seen, the probability of false alarm is particularly sensitive to the beam position, and we need to estimate the beam position accurately enough in order to minimize the number of false alarm events.}
  
%\subsection{Coarse estimate of receiver position}
%It may be the case that we may detect the presence of the detector array at one or more points on the spiral when we scan the entire uncertainty region. When that is the case, a coarse estimate of the location $(r_0, \theta_0)$ of the detector array center, denoted by $(\hat{r}_c, \hat{\theta}_c)$, is the point on the spiral that obtains the maximum photon count during the observation interval $T_{\text{obs}}$ for all the points $\ell_d$ where the receiver array was detected\footnote{We can write any point $\ell$ on the spiral in terms of polar coordinates as $(r, \theta)$. }:
%\begin{align}
%(\hat{r}_c, \hat{\theta}_c) = \argmax_{\ell_d} \sum_{m=1}^{M}z_m^{(\ell_d)}.
%\end{align}
%The first stage of acquisition is terminated at the point once the detector decides $H_0$ after having decided $H_1$ once, or more than once in a succession. 

\subsection{Mean Acquisition Time}
The acquisition time is defined to be the total time it takes for the completion of the acquisition stage. In this regard, we assume that the acquisition time is approximately equal to the to time it takes for the initiating terminal A in order to detect the presence of terminal B (please refer to Section~\ref{acq_approach}). Once terminal B is detected, it does not take too long for terminal B to detect or discover terminal A as discussed in Section~\ref{acq_approach}. Similarly, the amount of time it takes to send and receive data on the control channel is also assumed to be negligible. 

The approximate time for one full scan of the uncertainty region, denoted by $T_s$, is
\begin{align}
T_s  \approx N_s T_d, \label{Ta}
\end{align}
where $T_d$ is the dwell time. The mean acquisition time is defined as the time it takes from the start of the scanning process to the point where the receiver array is detected. It is modeled as a random variable and denoted by $T_a$. The acquisition time can be bounded from above as
\begin{align}
  T_a < T_U \triangleq T_s X +  T_d W, \label{bound}
\end{align}
where $X$ is the number of ``failed'' attempts (or the number of complete scans conducted) before the receiver array is detected, and $W$ is the additional time incurred during the last (successful) scan until the location of the array is discovered. The quantity $X$ is  \emph{geometric} random variable with probability of failure $p$. The probability 
\begin{align} 
p \triangleq P_m^{\lfloor N_0 - L/ \rho(z)\rfloor}, \label{p}
\end{align}
 which is actually the upper bound on $P_M$ (see \eqref{PM}). We use the convention that $X$ takes on integer values starting from 0. Then,
\begin{align}
\E[X] = \frac{p}{1-p} = \frac{P_m^{\left\lfloor N_0 - \frac{L}{\rho(z)} \right\rfloor} }{1- P_m^{\left\lfloor N_0 - \frac{L}{\rho(z)} \right\rfloor}} \approx \frac{F_{Z_0}\left( \frac{\mu_s}{\sigma^2_s} \gamma_0 \right)^{\left\lfloor N_0 - \frac{L}{\rho(z)} \right\rfloor} }{1- F_{Z_0}\left( \frac{\mu_s}{\sigma^2_s} \gamma_0 \right)^{\left\lfloor N_0 - \frac{L}{\rho(z)} \right\rfloor}}, \label{poiss}
\end{align}
where $Z_0 \sim$ Poisson$\left(\frac{\mu_s}{\sigma^2_s} \mu_s \right)$, and $F_{Z_0}(z)$ is the cumulative distribution function (cdf) of $Z_0$.  Substituting the expression of the cdf of the Poisson random variable into \eqref{poiss}, we have that
\begin{align}
\E[X] \approx  \frac{  \left(  \Gamma\left( \left\lfloor \frac{\mu_s}{\sigma_s^2} \gamma_0+1   \right\rfloor, \frac{\mu_s}{\sigma_s^2} \mu_s  \right) /  \left\lfloor \frac{\mu_s}{\sigma_s^2} \gamma_0 \right\rfloor ! \right)^{\left\lfloor N_0 - \frac{L}{\rho(z)} \right\rfloor} } {1 - \left(  \Gamma\left( \left\lfloor \frac{\mu_s}{\sigma_s^2} \gamma_0+1   \right\rfloor, \frac{\mu_s}{\sigma_s^2} \mu_s  \right) /  \left\lfloor \frac{\mu_s}{\sigma_s^2} \gamma_0 \right\rfloor ! \right)^{\left\lfloor N_0 - \frac{L}{\rho(z)} \right\rfloor}  }, \quad \gamma_0 \geq 0.
\end{align}

From \eqref{bound}, we can compute an upper bound on the expected acquisition time as
\begin{align}
\mu_{T_a} < \E[T_U] = T_s \E[X] +  T_d \E[W], \label{ubound}
\end{align}
where $\mu_{T_a} \triangleq \E[T_a]$.

In order to find the statistics of random variable $W$, let the distance of receiver's center be $R$ from the center of the uncertainty region.  Moreover, let $W$ be a random variable such that $
W \triangleq  \frac{\pi R^2}{\pi \rho^2(z)} = \frac{R^2}{\rho^2(z)}.$
Then $W$ represents an upper bound on the number of steps taken from the center of the spiral to the receiver location. Following the arguments in Section~\ref{acq_uncertain}, we see that  $R$ is a Rayleigh random variable (with some parameter $\sigma_0^2$):
$
f_R(r)  \triangleq \frac{r}{\sigma_0^2} \exp\left( {-\frac{r^2}{2\sigma_0^2}} \right) \cdot \mathbbm{1}_{[0, \infty)}(r).
$
  It can be shown that the density function of $W$ is $
f_W(w) = \frac{\rho^2(z)}{2 \sigma^2_0} \exp\left( -\frac{ \rho^2(z) }{2 \sigma_0^2}w   \right) \cdot \mathbbm{1}_{[0, \infty)}(w),$
where $W$ is an \emph{exponential} random variable with mean
 $ \displaystyle
\E[W] = \frac{2 \sigma_0^2}{\rho^2(z)}. \label{meanZ}
$
Furthermore, the upper bound in \eqref{poiss} can be loosened to admit a more tractable objective function (objective function being the upper bound on mean acquisition time) in terms of $\rho(z)$ as
\begin{align}
\E[X] < \frac{F_{Z_0}\left( \frac{\mu_s}{\sigma^2_s} \gamma_0 \right)^{ N_0 - \frac{L}{\rho(z)} -1 } }{1- F_{Z_0}\left( \frac{\mu_s}{\sigma^2_s} \gamma_0 \right)^{ N_0 - \frac{L}{\rho(z)} -1}} <       \frac{  \left[  \Gamma\left( \frac{\mu_s}{\sigma_s^2} \gamma_0+1  , \frac{\mu_s}{\sigma_s^2} \mu_s  \right) /  \Gamma\left( \frac{\mu_s}{\sigma_s^2} \gamma_0 + 1 \right) \right]^{ N_0 - \frac{L}{\rho(z)} -1} } {1 - \left[  \Gamma\left( \frac{\mu_s}{\sigma_s^2} \gamma_0+1  , \frac{\mu_s}{\sigma_s^2} \mu_s  \right) /  \Gamma\left( \frac{\mu_s}{\sigma_s^2} \gamma_0 + 1 \right) \right]^{ N_0 - \frac{L}{\rho(z)} -1}  }. \label{poiss1}
\end{align} 
Then, 
\begin{align}
\mu_{T_a} &< T_s \frac{  \left[  \Gamma\left( \frac{\mu_s}{\sigma_s^2} \gamma_0+1  , \frac{\mu_s}{\sigma_s^2} \mu_s  \right) /  \Gamma\left( \frac{\mu_s}{\sigma_s^2} \gamma_0 + 1 \right) \right]^{ N_0 - \frac{L}{\rho(z)} -1} } {1 - \left[  \Gamma\left( \frac{\mu_s}{\sigma_s^2} \gamma_0+1  , \frac{\mu_s}{\sigma_s^2} \mu_s  \right) /  \Gamma\left( \frac{\mu_s}{\sigma_s^2} \gamma_0 + 1 \right) \right]^{ N_0 - \frac{L}{\rho(z)} -1}  } + T_d \frac{2 \sigma_0^2}{\rho^2(z)} \label{ubound3} \\
& < \frac{R_u^2}{\rho^2(z)} T_d \times \frac{  \left[  \Gamma\left( \frac{\mu_s}{\sigma_s^2} \gamma_0+1  , \frac{\mu_s}{\sigma_s^2} \mu_s  \right) /  \Gamma\left( \frac{\mu_s}{\sigma_s^2} \gamma_0 + 1 \right) \right]^{ \frac{|\mathcal{A}|}{4\rho^2(z)} - \frac{L}{\rho(z)} -2} } {1 - \left[  \Gamma\left( \frac{\mu_s}{\sigma_s^2} \gamma_0+1  , \frac{\mu_s}{\sigma_s^2} \mu_s  \right) /  \Gamma\left( \frac{\mu_s}{\sigma_s^2} \gamma_0 + 1 \right) \right]^{ \frac{|\mathcal{A}|}{4\rho^2(z)} - \frac{L}{\rho(z)} -2}  } + T_d \frac{2 \sigma_0^2}{\rho^2(z)} \label{ubound2}
\end{align}
where the second inequality in \eqref{ubound2} is true because $N_0 \geq \frac{|\mathcal{A}|}{4\rho^2(z)} - 1$ and $N_s < \frac{R_u^2}{\rho^2(z)}$. For the sake of compact notation, let us denote the upper bound on $\E[T_a]$ in \eqref{ubound2} as $\mu_{T_a}^{(U)}$:
\begin{align}
 \mu_{T_a}^{(U)} \triangleq \frac{R_u^2}{\rho^2(z)} T_d \times \frac{  \left[  \Gamma\left( \frac{\mu_s}{\sigma_s^2} \gamma_0+1  , \frac{\mu_s}{\sigma_s^2} \mu_s  \right) /  \Gamma\left( \frac{\mu_s}{\sigma_s^2} \gamma_0 + 1 \right) \right]^{ \frac{|\mathcal{A}|}{4\rho^2(z)} - \frac{L}{\rho(z)} -2} } {1 - \left[  \Gamma\left( \frac{\mu_s}{\sigma_s^2} \gamma_0+1  , \frac{\mu_s}{\sigma_s^2} \mu_s  \right) /  \Gamma\left( \frac{\mu_s}{\sigma_s^2} \gamma_0 + 1 \right) \right]^{ \frac{|\mathcal{A}|}{4\rho^2(z)} - \frac{L}{\rho(z)} -2}  } + T_d \frac{2 \sigma_0^2}{\rho^2(z)} \label{Tu}.
\end{align}
Finally, it should be noted that $\E[T_U] < \mu_{T_a}^{(U)}$.

\subsection{Effect of Beam Radius on Mean Acquisition Time}
In this section, we analyze the effect of the beam radius $\rho(z)$ on the upper bound on mean acquisition time $\mu_{T_a}^{(U)}$. {The beam radius can be varied by using lens of different diameters in the transmitter telescope.  For large $z$ (large distance from the transmitter), a larger diameter of the telescope lens results in a larger beam radius $\rho(z)$ at the receiver, and vice versa \cite{Alda}.}

A smaller beam radius $\rho(z)$ implies a bigger likelihood of concentrating all the energy in the beam on the detector array leading to a smaller geometric loss. Moreover, for the same SNR, a smaller $\rho(z)$ also implies that the SNR is also high in those cells which are illuminated by the beam. Both these factors help in the minimization of the probability of missed detection, and it is likely that we may detect the array during the first scan. However, if the beam radius is too small, it may result in a larger scan time $T_s$, even though the probability of missed detection is smaller in this case. On the other hand, a bigger $\rho(z)$ helps in minimization of $T_s$ but results in a larger probability of missed detection. Hence, we have to optimize the average acquisition time as a function of $\rho(z)$.

Since an exact expression for mean acquisition time $\E[T_a]$ is not easy to compute, we can optimize the upper bound on  $\E[T_a]$ (bound is given in \eqref{ubound2}) with respect to $\rho(z)$ assuming that the signal power $S$ is fixed at some positive number $S_0$, and the upper bound on the probability of false alarm is less than some $P_0$:
\begin{equation*}
\begin{aligned}
& \underset{\rho(z)}{\text{minimize}}
& &  \mu_{T_a}^{(U)} \\
& \text{subject to}
& & \rho_{\text{min}} < \rho(z) < \rho_{\text{max}},\\
& & & S = S_0.
\end{aligned}
\end{equation*}
The quantities $\mu_s$ and $\sigma^2_s$ in \eqref{ubound2} are now explicitly defined in terms of $\rho(z)$:
\begin{align}
\mu_{s}= \sum_{m=1}^M \ln\left( 1 + \frac{\iint_{A_m} \frac{I_0}{\rho^2(z) } \exp\left(- \frac{(x-x_0)^2 + (y-y_0)^2}{2 \rho^2(z)}\right)\,dx \, dy    }{ \lambda_n |A_m| }  \right) K \iint_{A_m} \left( \frac{I_0}{\rho^2(z) } e^{- \frac{(x-x_0)^2 + (y-y_0)^2}{2 \rho^2(z)}}+ \lambda_n \right) \, dx \, dy, \label{mean}
\end{align}
\begin{align}
\sigma^2_{s}= \sum_{m=1}^M \left[ \ln\left( 1 + \frac{\iint_{A_m} \frac{I_0}{\rho^2(z) } \exp\left(- \frac{(x-x_0)^2 + (y-y_0)^2}{2 \rho^2(z)} \right)\, dx \, dy    }{ \lambda_n |A_m| }  \right) \right]^2 \! \! \! \!K \iint_{A_m} \left( \frac{I_0}{\rho^2(z) } e^{- \frac{(x-x_0)^2 + (y-y_0)^2}{2 \rho^2(z)}}+ \lambda_n \right) \, dx \, dy. \label{var}
\end{align}
In \eqref{mean} and \eqref{var}, we have scaled $I_0$ by $\rho(z)$ so that the received signal power is held constant regardless of the beam radius $\rho(z)$.

\begin{figure}
	\centering
	\includegraphics[scale=0.75]{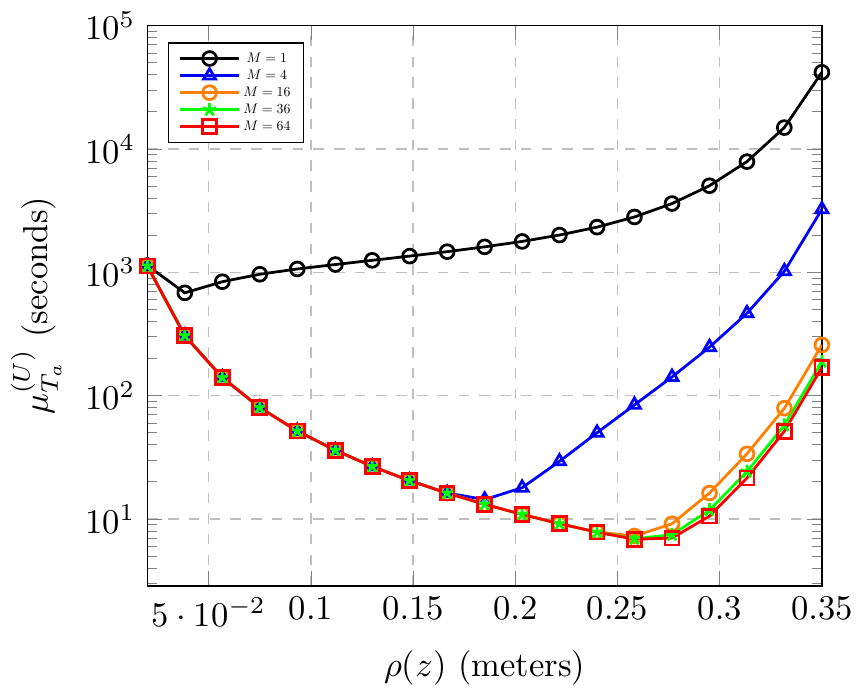}
	\caption{Plot of $T_U$ as a function of $\rho(z)$ for different detector arrays. The signal power and noise power was fixed at 1 $\mu$W, and $P_{F_U} < 7 \times 10^{-10}$.} \label{obj_f}
\end{figure}

\begin{figure}
	\centering
	\begin{subfigure}[b]{0.35\textwidth}
		\includegraphics[width=\textwidth]{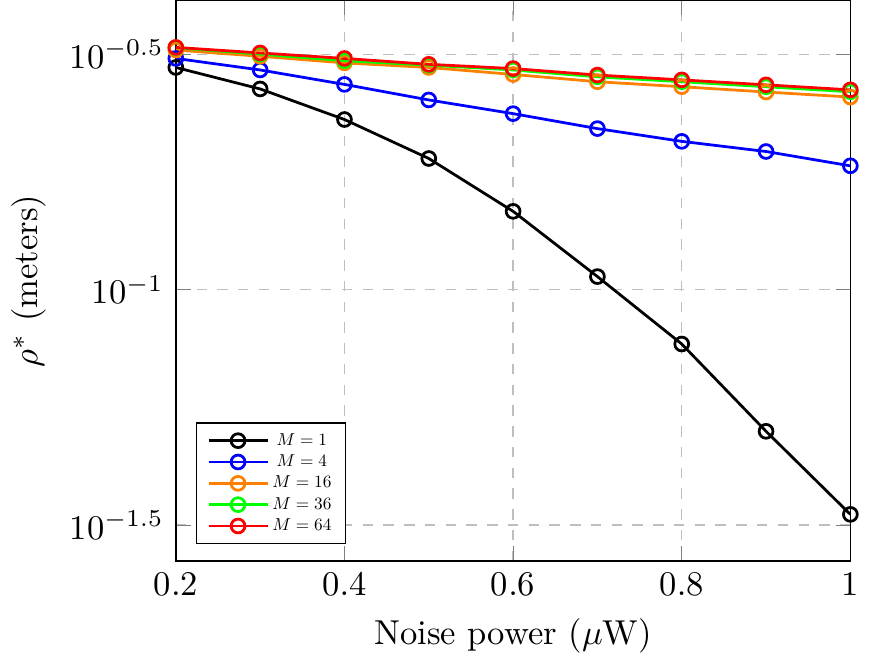}
		\caption{Plot 1}
		\label{fig3}
	\end{subfigure}
	\begin{subfigure}[b]{0.35\textwidth}
		\includegraphics[width=\textwidth]{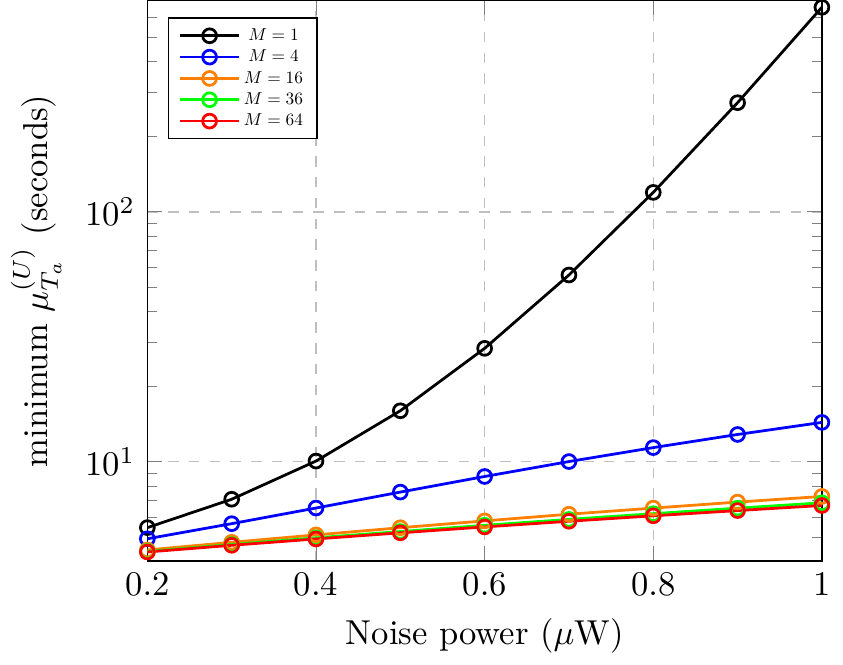}
		\caption{Plot 2}
		\label{fig4}
	\end{subfigure}
\caption{Fig.~\ref{fig3} depicts $\rho^*(z)$ as a function of noise power, and Fig.~\ref{fig4} shows the minimum value of the upper bound on acquisition time as a function of noise power. The signal power was held constant at 1 $\mu$W, $P_{F_U} < 7\times 10^{-10}$, and $T_d = 1$ millisecond. } \label{fig_optim}
\end{figure}

Fig.~\ref{obj_f} depicts the objective function $\mu_{T_a}^{(U)}$ (defined in \eqref{Tu}) as a function of $\rho(z)$ for different types of detector arrays. Because $\mu_{T_a}^{(U)}$ is not a tractable function in terms of $\rho(z)$, we resort to numerical optimization techniques for minimizing $\mu_{T_a}^{(U)}$ as a function of $\rho(z)$.

Fig.~\ref{fig_optim} shows the curves that indicate the optimum values of $\rho(z)$ and the corresponding minimum upper bound on the mean acquisition time achieved with the optimum $\rho(z)$.  If we analyze $\mu_{T_a}^{(U)}$ in \eqref{Tu}, we find that it is made up of three terms: the first term corresponds to $T_s$, the following term (containing upper incomplete gamma function terms in the numerator and denominator) represents the probability of missed detection, and the last term that represents the mean value of acquisition time only for the (last) successful scan. The first and the third terms decrease monotonically with $\rho(z)$, whereas the second term increases monotonically with $\rho(z)$ (since the probability of missed detection goes up if the beam energy is spread out for the constant signal power). When $\rho(z)$ is very small ($\rho(z) \to 0$), the dominant source of delay is the first and third terms, and when $\rho(z)$ is large, then the second term represents the main contributing factor to a larger delay in the acquisition process. For $\rho(z) \to \infty$, the third term goes to zero, and the product of the first and second term goes to infinity (because the second term is approaching infinity faster than the first term approaching zero).

{Fig.~\ref{fig3} represents the effect on the optimum value $\rho^*(z)$ as a function of noise power, and Fig.~\ref{fig4} depicts the relationship betwen the minimum value of the upper bound on acquisition time and the noise power.} We can see that the the optimum $\rho^*(z)$ decreases monotonically with the noise power. For a small noise power, the probability of missed detection is approximately small for all value of $\rho(z)$. Therefore, we pick a relatively larger $\rho(z)$ in order to minimize the overall time delay by minimizing the first and the third terms. The same argument also explains why $\rho^*(z)$ is large for higher order detector arrays since the probability of missed detection is better for such detector arrays. Fig.~\ref{fig4} represents the upper bound on mean acquisition time as a function of noise power for different detector arrays. 

In Fig.~\ref{fig3}, $\rho^*(z)$ was found over the interval $(0, 0.35)$ since a beam of radius 0.35 m covers the entire detector array with its footprint. Moreover, for $\rho(z) > 0.35$, the exponent term $\frac{|\mathcal{A}|}{4\rho^2(z)} - \frac{L}{\rho(z)} -2$ in \eqref{Tu} becomes less than 1 which is not a practical representation.

\subsection{Complementary Distribution Function of $T_U$}
We are interested in determining the complementary distribution function of $T_U$:
\begin{align}
P(\{T_U \geq \gamma\}) = P(\{ T_s X + T_d W  \geq \gamma \}), \quad \gamma > 0.
\end{align}
Let $Y \triangleq T_s X$ and $V \triangleq T_d W$. Then, for a nonnegative integer $k$
\begin{align}
P(\{Y = k T_s\}) &= P(\{ T_s X = k T_s  \}) = P(\{ X = k  \}=p^k (1-p),
\end{align}
where the probability of failure $p$ is defined in \eqref{p}.
Moreover, in terms of the probability density function $f_Y(y)$,
\begin{align}
f_Y(y) = \sum_{k=0}^{\infty} p^k(1-p)\delta(y-kT_s),
\end{align}
where $\delta(x)$ is the Dirac delta function.
Moreover, it can be shown easily that 
\begin{align}
f_V(v) = \frac{\rho^2(z)}{2T_d  \sigma_0^2} \exp\left( -\frac{\rho^2(z)}{2T_d \sigma_0^2} v  \right) \cdot \mathbbm{1}_{(0, \infty)}(v)
\end{align}
so that $V$ is an exponential random variable with mean $\dfrac{2 T_d \sigma_0^2}{\rho^2(z)}$. In order to minimize space, let us define $\beta \triangleq \frac{\rho^2(z)}{2 T_d \sigma_0^2}$. Then, 
\begin{align}
f_{T_U}(t) &f_{T_U}(t) = f_Y(t) * f_V(t)
= \beta \exp\left( -\beta t  \right) \cdot \mathbbm{1}_{(0, \infty)}(t) * \sum_{k=0}^{\infty} p^k(1-p)\delta(t-kT_s) \nonumber \\
&= \beta (1-p)\sum_{k=0}^{\infty} p^k \exp\left( -\beta (t - k T_s)  \right) \cdot \mathbbm{1}_{(kT_s, \infty)}(t). \label{exp}
\end{align}
\begin{figure}
	\centering
	\hspace{-1cm}	\begin{subfigure}{0.4\textwidth}
		\includegraphics[width=\textwidth]{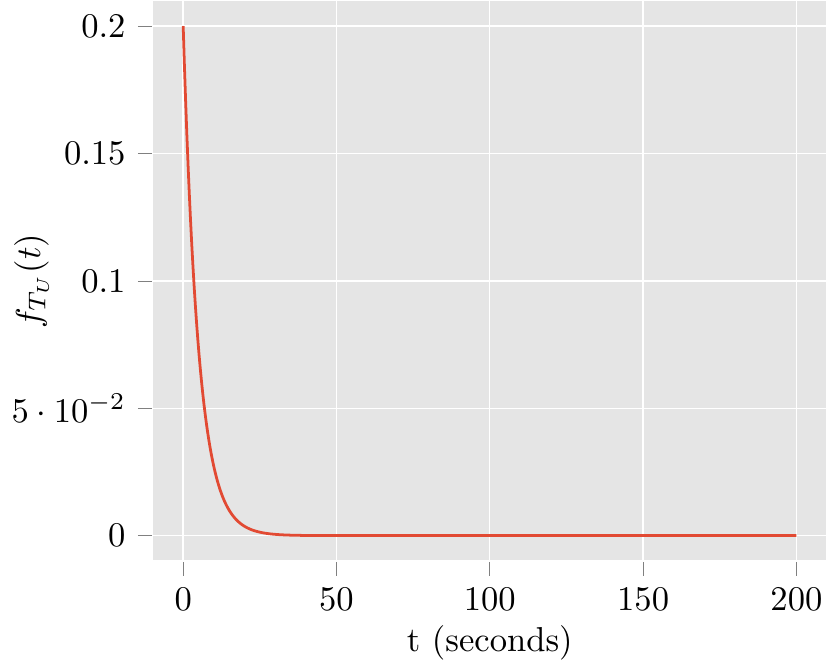}
		\caption{Plot (a)} \label{exp_plot}
	\end{subfigure} 
	\begin{subfigure}{0.4\textwidth}
		\includegraphics[width=\textwidth]{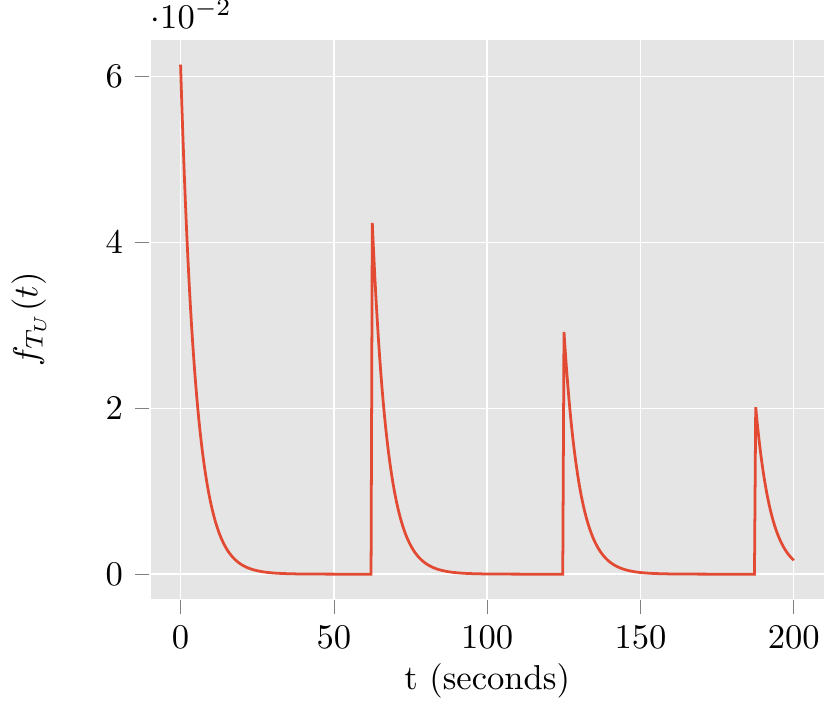}
		\caption{Plot (b)} \label{comb_plot}
	\end{subfigure}
\vspace{-0.3cm}
	\caption{The density function of $T_U$ as a function of time. In Fig.~\ref{exp_plot}, the signal power is 1 $\mu$W, whereas in Fig.~\ref{comb_plot}, the signal power is $7\times 10^{-7}$ Watts. The noise power is $1 \times 10^{-6}$ Watts in both cases. The beam radius $\rho(z) = 0.2$ meters, $R_u = 50$ meters, $\sigma_0$ = 10 meters, and $M=16$.} \label{pdf}
\end{figure}

\begin{figure}
	\centering
	\hspace{-1cm}\begin{subfigure}{0.4\textwidth}
		\includegraphics[width=\textwidth]{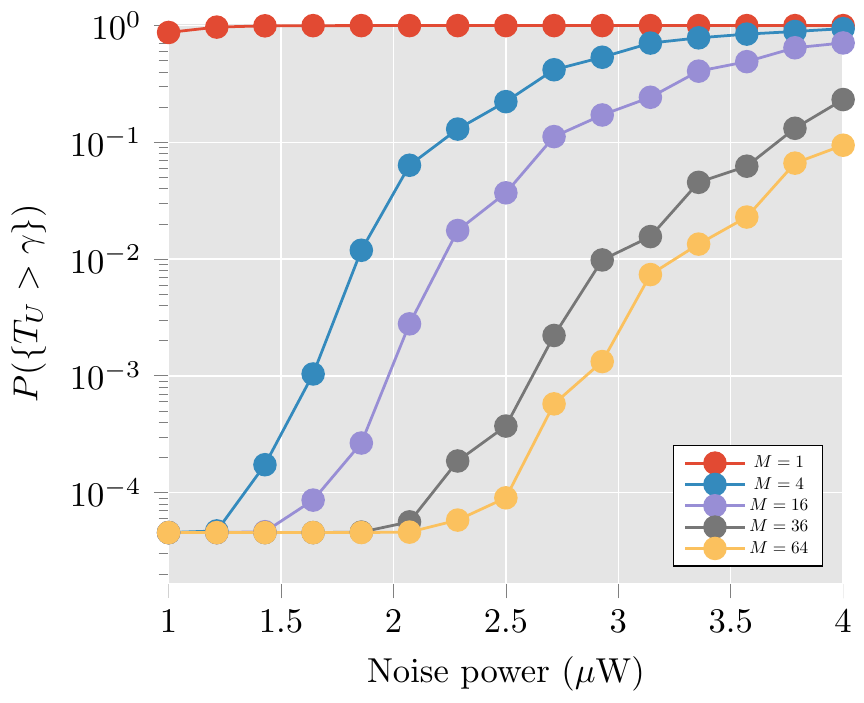} 
		\caption{Plot (a)} \label{noise}
	\end{subfigure}
	\vspace{-3cm}
	\begin{subfigure}{0.4\textwidth}
		\includegraphics[width=\textwidth]{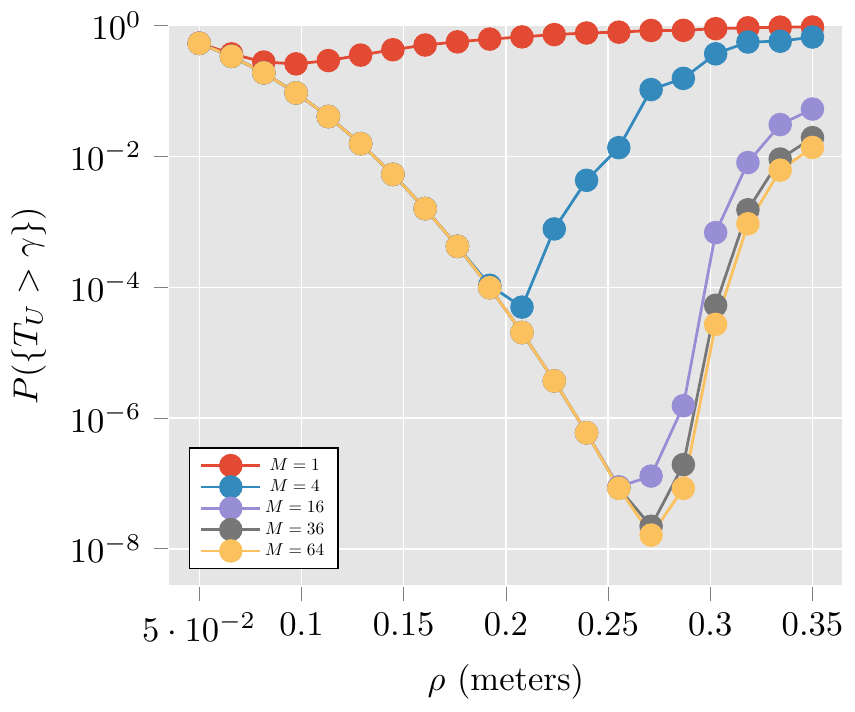} 
		\caption{Plot (b)} \label{p1}
	\end{subfigure}
	\vspace{2.9cm}
	\caption{Plot of the tail probability of $T_U$ as a function of noise power and beam radius for different detector arrays. The signal power is 1 $\mu$W, $\rho(z)=0.2$ m, $R_u= 50$ meters, $\sigma_0 = 10$ meters, and $\gamma = 50$ seconds for Fig.~\ref{noise}. For Fig.~\ref{p1}, the same values hold as for Fig.~\ref{noise} except that the noise power is fixed at 1 $\mu$W and $\rho(z)$ is varied.} \label{ccdf}
\end{figure}
where $*$ represents the convolution operator. Then, for any real $\gamma > 0$,
\begin{align}
&P(\{ T_U > \gamma\}) = \int_{\gamma}^{\infty} f_{T_U}(t) \, dt = \beta (1-p)\sum_{k=0}^{\infty} p^k \int_{\gamma}^{\infty} \exp\left( -\beta (t - k T_s)  \right) \cdot \mathbbm{1}_{(kT_s, \infty)}(t)\, dt \nonumber \\
&= \beta (1-p)\sum_{k=0}^{\infty} p^k \int_{\max(\gamma, k T_s)}^{\infty} \exp\left( -\beta (t - k T_s)  \right) \, dt = (1-p)\sum_{k=0}^{\infty} p^k  \exp\left( -\beta (\max(\gamma, kT_s) - kT_s )  \right) \nonumber \\
&= (1-p) \left( \sum_{k=0}^{\left\lfloor \frac{\gamma}{T_s} \right\rfloor}p^k \exp\left( - \beta (\gamma-kT_s) \right) + \sum_{k= \left\lceil \frac{\gamma}{T_s} \right\rceil}^{\infty} p^k    \right)= (1-p)\left( \exp\left( -\beta \gamma \right) \sum_{k=0}^{\left\lfloor \frac{\gamma}{T_s} \right\rfloor} p^k \exp\left( \beta kT_s  \right) +  \sum_{k= \left\lceil \frac{\gamma}{T_s} \right\rceil}^{\infty} p^k\right)    \nonumber\\
&= (1-p) \left( \exp\left( -\beta \gamma\right)  \sum_{k=0}^{\left\lfloor \frac{\gamma}{T_s} \right\rfloor} \left(p \exp \left( \beta T_s \right) \right)^k +  \left( \sum_{k=0}^{\infty} p^k - \sum_{k=0}^{\left\lfloor \frac{\gamma}{T_s} \right\rfloor} p^k \right)  \right) \nonumber \\
&= (1-p) \left( \exp\left( -\beta \gamma \right)  \times \frac{1-\left(p \exp \left( \beta T_s  \right) \right)^{\left\lceil \frac{\gamma}{T_s} \right\rceil}}{1-p \exp\left( \beta T_s  \right)  } + \frac{p^{\left\lceil \frac{\gamma}{T_s}  \right\rceil}}{1-p}  \right). 
\end{align}
Fig.~\ref{pdf} represents the probability density function of $T_U$ for the high/low SNR scenarios. For high SNR, the possibility of multiple scans is small, and the probability density function is approximately an exponential distribution with mean $\dfrac{1}{\beta}$. This can be seen by setting $p=0$ in \eqref{exp}. For low SNR, $p \approx 1$, and this results in a weighted train of exponential distributions corresponding to the possibility of multiple scans as shown in Fig.~\ref{comb_plot}.

Fig.~\ref{ccdf} shows the plots of the complementary cumulative distribution function of $T_U$ as a function of noise power and beam radius $\rho(z)$. This figure indicates the improvement in acquisition time that comes with higher order detector arrays. 

\section{A Brief Complexity Analysis} \label{complexity}
{The computational complexity at each step of the scan is given by the computation of the likelihood decision given by \eqref{detect}. In this regard, let $N \triangleq \sqrt{M}$ denote the number of cells on one side of the square detector array. We note that the computational complexity of the left hand side of \eqref{detect} is a function of the number of detectors $N^2$ of the array. Thus, assuming that a lookup table containing all values of the Gaussian distribution function is available, we see that there are roughly $3N^2$ real additions and subtractions, and $3N^2$ real multiplications. Thus, the complexity grows as $N^2$ as a function of $N$. This is in addition to the complexity of finding the distribution values from the lookup table for each of the $N^2$ detectors of the array.}

{In addition to the computational complexity, there is also a hardware complexity associated with higher order detector arrays. A large number of detectors lead to a more complex Readout Integrated Circuit (ROIC) design, especially for significantly faster data-rates. Furthermore, for a large number of elements in the array, custom microlens arrays are required for increased optical coupling in order to compensate for the reduction in the optical fill factors \cite{Bashir4}. Finally, the storage complexity also goes up as a factor of $N^2$ (we need to store $N^2$ photon count values in order to compute the sufficient statistic for the evaluation of \eqref{detect}).}

\section{Conclusions} \label{conc}

In this paper, we have examined the acquisition problem in free-space optical communications with detector arrays.  {Based on the results of Fig.~\ref{obj_f} and Fig.~\ref{ccdf}, we conclude that an array of smaller detectors is  more advantageous from the acquisition perspective as compared to a single detector of similar dimensions. However, as can be seen from the same figures, the difference in improvement in acquisition performance shrinks as the number of detectors is increased (law of diminishing returns). Additionally, this improvement comes with the overhead of  extra computational and hardware complexity which is a function of the number of detectors in the array.}
	\bibliography{reference1}

\begin{thebibliography}{10}
\providecommand{\url}[1]{#1}
\csname url@samestyle\endcsname
\providecommand{\newblock}{\relax}
\providecommand{\bibinfo}[2]{#2}
\providecommand{\BIBentrySTDinterwordspacing}{\spaceskip=0pt\relax}
\providecommand{\BIBentryALTinterwordstretchfactor}{4}
\providecommand{\BIBentryALTinterwordspacing}{\spaceskip=\fontdimen2\font plus
\BIBentryALTinterwordstretchfactor\fontdimen3\font minus
  \fontdimen4\font\relax}
\providecommand{\BIBforeignlanguage}[2]{{%
\expandafter\ifx\csname l@#1\endcsname\relax
\typeout{** WARNING: IEEEtran.bst: No hyphenation pattern has been}%
\typeout{** loaded for the language `#1'. Using the pattern for}%
\typeout{** the default language instead.}%
\else
\language=\csname l@#1\endcsname
\fi
#2}}
\providecommand{\BIBdecl}{\relax}
\BIBdecl

\bibitem{Li}
L.~Li, R.~Zhang, Z.~Zhao, G.~Xie, P.~Liao, K.~Pang, H.~Song, C.~Liu, Y.~Ren,
  G.~Labroille, P.~Jian, D.~Starodubov, B.~Lynn, R.~Bock, M.~Tur, and A.~E.
  Willner, ``High-capacity free-space optical communications between a ground
  transmitter and a ground receiver via a {UAV} using multiplexing of multiple
  orbital-angular-momentum beams,'' \emph{Scientific Reports}, vol.~7, no.
  17427, December 2017.

\bibitem{Kaymak}
Y.~Kaymak, R.~Rojas-Cessa, J.~Feng, N.~Ansari, M.~Zhou, and T.~Zhang, ``A
  survey on acquisition, tracking and pointing mechanisms for mobile free-space
  optical communications,'' \emph{IEEE Communications Surveys \& Tutorials},
  vol.~20, no.~2, Second Quarter 2018.

\bibitem{Ansari}
I.~Ansari, F.~Yilmaz, and {M.~-S.~Alouini}, ``Performance analysis of
  free-space optical links over {M}alaga ({M}) turbulence channels with
  pointing errors,'' \emph{IEEE Transactions on Wireless Communications},
  vol.~15, no.~1, pp. 91--102, January 2016.

\bibitem{Zedini}
E.~Zedini, H.~Soury, and {M.~-S.~Alouini}, ``Dual-hop {FSO} transmission
  systems over {Gamma-Gamma} turbulence with pointing errors,'' \emph{IEEE
  Transactions on Wireless Communications}, vol.~16, no.~2, pp. 784--796,
  February 2017.

\bibitem{Quwaiee}
H.~Al-Quwaiee, H.~C. Yang, and {M.~-S.~Alouini}, ``On the asymptotic capacity
  of dual-aperture {FSO} systems with a generalized pointing error model,''
  \emph{IEEE Transactions on Wireless Communications}, vol.~15, no.~9, pp.
  6502--6512, September 2016.

\bibitem{issaid_TWC_2017}
C.~B. Issaid, K.-H. Park, and M.-S. Alouini, ``A generic simulation approach
  for the fast and accurate estimation of the outage probability of single hop
  and multihop {FSO} links subject to generalized pointing errors,''
  \emph{{IEEE} Trans. Wireless Commun.}, vol.~16, no.~10, pp. 6822--6837, 2017.

\bibitem{Wang}
J.~Wang, J.~M. Kahn, and K.~Y. Lau, ``Minimization of acquisition time in
  short-range free-space optical communication,'' \emph{Applied Optics},
  vol.~41, no.~36, December 2002.

\bibitem{XinLi}
X.~Li, S.~Yu, J.~Ma, and L.~Tan, ``Analytical expression and optimization of
  spatial acquisition for intersatellite optical communications,'' \emph{Optics
  Express}, vol.~19, no.~3, pp. 2381--2390, January 2011.

\bibitem{Farid}
A.~A. Farid and S.~Hranilovic, ``Outage capacity optimization for free-space
  optical links with pointing errors,'' \emph{Journal of Lightwave Technology},
  vol.~25, no.~7, July 2007.

\bibitem{Mai}
V.~V. Mai and H.~Kim, ``Adaptive beam control techniques for airborne
  free-space optical communication systems,'' \emph{Applied Optics}, vol.~57,
  no.~26, September 2018.

\bibitem{Deng}
P.~Deng, T.~Kane, and O.~Alharbi, ``Reconfigurable free space optical data
  center network using gimbal-less {MEMS} retroreflective acquisition and
  tracking,'' in \emph{Proc. {SPIE} 10524 Free-Space Laser Communication and
  Atmospheric Propagation ({SPIE LASE}' 18)}, San Francisco, California, United
  States, Feb. 2018.

\bibitem{Kim}
J.~J. Kim, T.~Sands, and B.~N. Agrawal, ``Acquisition, tracking, and pointing
  technology development for bifocal relay mirror spacecraft,'' in \emph{Proc.
  SPIE vol.~6569, Acquisition, Tracking, Pointing, and Laser Systems
  Technologies XXI}, Orland, Florida, United States, May 2007.

\bibitem{Rzasa}
J.~Rzasa, M.~C. Ertem, and C.~C. Davis, ``Pointing, acquisition, and tracking
  considerations for mobile directional wireless communications systems,'' in
  \emph{Proc. SPIE 8874, Laser Communication and Propagation through the
  Atmosphere and Oceans II, 88740C}, San~Diego California United States, 2013.

\bibitem{Kaushal1}
H.~Kaushal, G.~Kaddoum, V.~K. Jain, and S.~Kar, ``Experimental investigation of
  optimum beam size for fso uplink,'' \emph{Optics Communications}, vol. 400,
  pp. 106--114, October 2017.

\bibitem{Cova}
S.~D. Cova and M.~Ghioni, ``Single-photon counting detectors,'' \emph{IEEE
  Photonics Journal}, vol.~3, no.~2, April 2011.

\bibitem{Bashir4}
M.~S. {Bashir}, ``Free-space optical communications with detector arrays: A
  mathematical analysis,'' \emph{IEEE Transactions on Aerospace and Electronic
  Systems}, pp. 1--1, 2019.

\bibitem{Bashir2}
M.~S. Bashir and M.~R. Bell, ``Optical beam position tracking in free-space
  optical communication systems,'' \emph{IEEE Transactions on Aerospace and
  Electronic Systems}, vol.~20, no.~2, April 2018.

\bibitem{Bashir3}
M.~S. {Bashir} and M.~R. {Bell}, ``The impact of optical beam position
  estimation on the probability of error in free-space optical
  communications,'' \emph{IEEE Transactions on Aerospace and Electronic
  Systems}, vol.~55, no.~3, pp. 1319--1333, June 2019.

\bibitem{Bashir1}
M.~S. Bashir and M.~R. Bell, ``Optical beam position estimation in free-space
  optical communication,'' \emph{IEEE Transactions on Aerospace and Electronic
  Systems}, vol.~52, no.~6, December 2016.

\bibitem{Snyder}
D.~L. Snyder and M.~I. Miller, \emph{Random Point Processes in Time and
  Space}.\hskip 1em plus 0.5em minus 0.4em\relax New York, NY: Springer-Verlag,
  1991.

\bibitem{Streit}
R.~L. Streit, \emph{Poisson Point Processes: Imaging, Tracking and
  Sensing}.\hskip 1em plus 0.5em minus 0.4em\relax New York, NY: Springer,
  2010.

\bibitem{Kaushal}
H.~Kaushal, V.~K. Jain, and S.~Kar, \emph{Free Space Optical
  Communications}.\hskip 1em plus 0.5em minus 0.4em\relax New Delhi, India:
  Springer, 2017.

\bibitem{Bloom}
S.~Bloom, E.~Korevaar, J.~Schuster, and H.~Willebrand, ``Understanding the
  performance of free-space optics,'' \emph{Journal of Optical Networking},
  vol.~2, no.~6, June 2003.

\bibitem{Tan}
Y.~Tan, Y.~Lu, and K.~H. Xia, ``Relative error of scaled poisson approximation
  via {Stein's} method,'' arXiv:1810.04300v1~[math.PR].

\bibitem{Alda}
J.~Alda, \emph{Encyclopedia of Optical Engineering}.\hskip 1em plus 0.5em minus
  0.4em\relax New York, NY: Marcel Dekker Inc., 2003.

\end{thebibliography}
	\bibliographystyle{IEEEtran}
	\vspace{-1cm}	\begin{IEEEbiography}
		[{\includegraphics[width=1in,height=1.25in,clip,keepaspectratio]{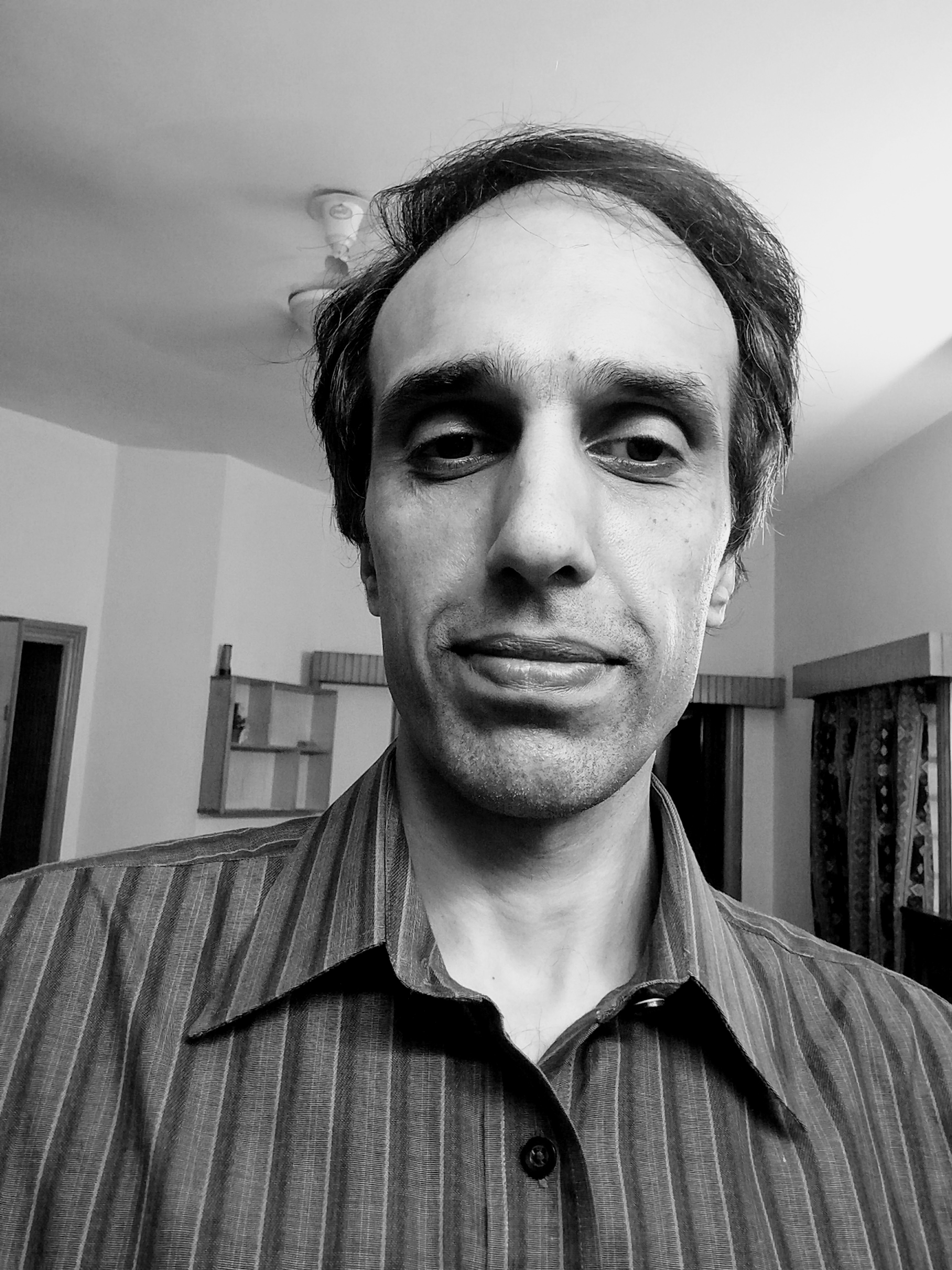}}]{Muhammad Salman Bashir} (S'14-M'17)
		received his M.S. and  Ph.D. degrees in Electrical and Computer Engineering from Purdue University, West Lafayette, IN, US, in 2014 and 2017, respectively. He was a recipient of International Fulbright Science and Technology award for his graduate studies at Purdue University. He served as a faculty member at the National University of Computer and Emerging Sciences Lahore from 2017 till 2018. Since 2019, he has been working as a Postdoctoral Fellow in the Communication Theory Lab at King Abdullah University of Science and Technology (KAUST). Salman is a member of IEEE, Golden Key and Eta Kappa Nu. His research interests are mainly in the application of estimation theory and target tracking to pointing, acquisition and tracking problems in free-space optical communications.
		
	\end{IEEEbiography}
	\vspace{-1cm}
	\begin{IEEEbiography} [{\includegraphics[width=1in,height=1.25in,clip,keepaspectratio]{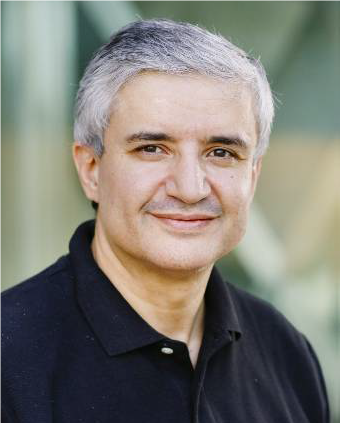}}]{Mohamed-Slim~Alouini} 	(S'94-M'98-SM'03-F'09)  was	born in Tunis, Tunisia. He received the Ph.D. degree in Electrical Engineering
	from the California Institute of Technology (Caltech), Pasadena, CA, USA, in 1998. He served as a faculty member in the University of Minnesota, Minneapolis, MN, USA, then in the Texas A\&M University at Qatar, Education City, Doha, Qatar before joining King Abdullah University of Science and Technology (KAUST), Thuwal, Makkah Province, Saudi Arabia as a Professor of Electrical Engineering in 2009. His current research interests include the modeling, design, and
	performance analysis of wireless communication systems.
\end{IEEEbiography}

\end{document}